\documentclass[english,showpacs,twocolumn,superscriptaddress,nofootinbib,aps]{revtex4-1}
\usepackage[T1]{fontenc}
\usepackage[latin9]{inputenc}
\usepackage{color}
\usepackage{babel}
\usepackage{float}
\usepackage{booktabs}
\usepackage{textcomp}
\usepackage{amsmath}
\usepackage{amssymb}
\usepackage{stmaryrd}
\usepackage{graphicx}
\usepackage[unicode=true,pdfusetitle,
 bookmarks=true,bookmarksnumbered=false,bookmarksopen=false,
 breaklinks=false,pdfborder={0 0 1},backref=false,colorlinks=true]
 {hyperref}

\makeatletter

\providecommand{\tabularnewline}{\\}

\makeatother

\begin{document}
\title{Collision centrality and energy dependence of strange hadron production
in Au + Au collisions at $\sqrt{s_{NN}}=$ 7.7-54.4 GeV}
\author{Yan-ting Feng}
\affiliation{School of Physics and Physical Engineering, Qufu Normal University,
Shandong 273165, China}
\author{Zi-yao Song}
\affiliation{School of Physics and Physical Engineering, Qufu Normal University,
Shandong 273165, China}
\author{Feng-lan Shao}
\email{shaofl@mail.sdu.edu.cn}

\affiliation{School of Physics and Physical Engineering, Qufu Normal University,
Shandong 273165, China}
\author{Jun Song }
\email{songjun2011@jnxy.edu.cn}

\affiliation{School of Physical Science and Intelligent Engineering, Jining University,
Shandong 273155, China}
\begin{abstract}
We apply an equal-velocity quark combination model to systematically
study the transverse momentum ($p_{T}$) spectra of strange hadrons
\textcolor{black}{$K_{S}^{0}$}, $\phi$, $\Lambda$, $\Xi^{-}$,
$\Omega^{-}$, $\bar{\Lambda}$, $\bar{\Xi}^{+}$ and $\bar{\Omega}^{+}$
at mid-rapidity in Au+Au collisions at $\sqrt{s_{NN}}=$ 7.7, 11.5,
19.6, 27, 39, 54.4 GeV. Relative deviation between the model calculation
and experimental data of these eight hadrons is generally about 2-3\%
at $\sqrt{s_{NN}}=$ 27, 39, 54.4 GeV and in central collisions at
7.7, 11.5, 19.6 GeV. The deviation slightly increases up to about
4\% in the semi-central and peripheral collision at $\sqrt{s_{NN}}=$
7.7, 11.5, 19.6 GeV. We systematically explain the dependence of two
baryon-to-meson ratios $\bar{\Lambda}/K_{S}^{0}$ and $\Omega/\phi$
on $p_{T}$, collision centrality and collision energy by the property
of quark $p_{T}$ spectra at hadronization. We derive the analytic
relations between $R_{CP}$ of hadrons and those of quarks, and we
use them to naturally explain the species and $p_{T}$ dependence
of $R_{CP}$ of those strange hadrons.
\end{abstract}
\maketitle

\section{Introduction\label{sec:Intro} }

Strange hadrons are excellent probes of high-energy collisions. The
enhancement of strange hadrons was proposed as a signal of quark-gluon
plasma (QGP) formation \citep{Rafelski:1982pu}, and it was widely
observed in relativistic heavy-ion collisions at SPS, RHIC and LHC
\citep{STAR:2007cqw,ALICE:2013xmt,NA49:2008ysv}. In recent years,
Beam Energy Scan (BES) experiments of STAR collaboration at RHIC have
obtained rich experimental data of yield and transverse momentum ($p_{T}$)
spectra of strange hadrons in Au+Au collisions at $\sqrt{s_{NN}}=$
7.7-54.4 GeV \citep{Adam:2019koz,Adamczyk:2015lvo,Muhammad:2019hbu,Huang:2021hbu,Huang:2022hbu}.
These experimental data contain valuable information on the property
of hot nuclear matter and QCD phase transition at the finite baryon
chemical potential, the mechanism of hadron production at hadronization,
etc. 

Existing theoretical studies\textbf{ }on these newest data of strange
hadrons at STAR are mainly of the global information on hadron freeze-out.
Statistical model analysis on yield data of strange hadrons obtain
the temperature and baryon chemical potential at the chemical freeze-out
of hadrons \citep{Adam:2019koz,Flor:2020fdw}. Analysis on $p_{T}$
spectra data of strange hadrons provide the kinematic freeze-out temperature
and collective radial flow at kinetic freeze-out of hadrons \citep{Adam:2019koz}.
There also exist phenomenological extension by introducing Tsallis
statistics at kinetic freeze-out to obtain a non-equilibrium parameter
$q$ besides the temperature and radial velocity \citep{Chen:2020zuw,Li:2022gxy,Waqas:2020ygr}.\textbf{ }

On the other hand, the microscopic mechanism of strange hadrons production
from the final-state parton systems created in heavy-ion collisions
at STAR BES energies is also necessary to be studied in details. For
example, we have known from the early heavy-ion collision experiments
at RHIC \citep{Greco:2003mm,Fries:2003vb,Molnar:2003ff,Hwa:2002tu,Chen:2006vc}
that the quark combination mechanism at hadronization is an effective
mechanism to explain the hadron production in relativistic heavy-ion
collisions. How about the performance of this mechanism at STAR BES
energies? The answer is not completely clear since the existing theoretical
studies and comparison with the newest STAR data are relatively lack
\citep{Jin:2018lbk,Song:2020kak,Ye:2017ooc}. Further studies are
necessary for deeply understanding the hadron production in heavy-ion
collisions at STAR BES energies, which can also serve as the basis
to better understand the hadron production in heavy-ion collisions
at lower energies in fixed target experiments of STAR collaboration.

\textbf{}

In this paper, we apply a quark combination model \citep{Song:2017gcz,Song:2020kak,Song:2018tpv}
to carry out a systematic study on $p_{T}$ spectra of strange hadrons
$K_{S}^{0}$, $\phi$, $\Lambda$, $\Xi^{-}$, $\Omega^{-}$, $\bar{\Lambda}$,
$\bar{\Xi}^{+}$ and $\bar{\Omega}^{+}$ at mid-rapidity in Au+Au
collisions in different centralities at $\sqrt{s_{NN}}=$ 7.7, 11.5,
19.6, 27, 39, 54.4 GeV. By a global fit to experimental data for $p_{T}$
spectra of eight hadrons, we study the performance of equal-velocity
combination (EVC) mechanism of constituent quark and antiquarks at
hadronization, and we study the significance of the hadronization
process imprinted in the final observation of strange hadrons. We
also study two baryon-to-meson ratios $\bar{\Lambda}/K_{S}^{0}$ and
$\Omega/\phi$ as the function of $p_{T}$, where we focus on the
self-consistent explanation on the collision centrality and energy
dependence of two ratios by the properties of quark $p_{T}$ distributions
at hadronization. In study of the nuclear modification factor $R_{CP}$
of strange hadrons, we drive several analytic relations between $R_{CP}$
of hadrons and those of quarks, and we use them to naturally explain
the species dependence of $R_{CP}$ of different hadrons. 

The paper is organized as follows. In Sec.~\ref{sec:EVC_model},
we briefly introduce the model we used in this paper. In Sec.~\ref{sec had_pt_spectra},
we show the global description of $p_{T}$ spectra of strange hadrons
in different centralities at $\sqrt{s_{NN}}=$ 7.7-54.4 GeV. In Sec.~\ref{sec:BARYON TO MESON RATIOS },
we discuss the centrality and collision energy dependence of baryon-to-meson
ratios. In Sec.~\ref{sec:Rcp}, we study the nuclear modification
factor of strange hadrons. The summary is given in Sec.~\ref{sec:Summary}. 

\section{A QUARK COMBINATION MODEL WITH EVC\label{sec:EVC_model}}

In this section, we briefly introduce a quark combination model used
in this paper. The model was firstly proposed in \citep{Song:2017gcz}
based on the finding of the quark number scaling property of hadronic
$p_{T}$ spectra in $p$Pb collisions at LHC energy. Subsequently,
this scaling property was further found in $pp$ and AA collisions
at both RHIC and LHC energies and the model was systematically tested
by the experimental data of hadronic $p_{T}$ spectra and elliptic
flow in those collisions \citep{Song:2018tpv,Li:2017zuj,Zhang:2018vyr,Song:2019sez,Song:2020kak,Li:2021nhq,Wang:2019fcg}. 

In the framework of quark combination mechanism, the inclusive momentum
distribution of baryon $\left(B_{j}\right)$ and meson $\left(M_{j}\right)$
can be obtained by

\begin{align}
f_{B_{j}}\left(p_{B}\right) & =\int dp_{1}dp_{2}dp_{3}f_{q_{1}q_{2}q_{3}}\left(p_{1},p_{2},p_{3}\right)\label{eq:fbi}\\
 & \,\,\,\,\times\mathcal{R}_{B_{j}}\left(p_{1},p_{2},p_{3};p_{B}\right),\nonumber \\
f_{M_{j}}\left(p_{M}\right) & =\int dp_{1}dp_{2}f_{q_{1}\bar{q}_{2}}\left(p_{1},p_{2}\right)\mathcal{R}_{M_{j}}\left(p_{1},p_{2};p_{M}\right).\label{eq:fmi}
\end{align}
Here, $f_{q_{1}q_{2}q_{3}}\left(p_{1},p_{2},p_{3}\right)$ is the
joint momentum distribution function for $q_{1},q_{2},q_{3}$ and
$f_{q_{1}\bar{q}_{2}}\left(p_{1},p_{2}\right)$ is that for $q_{1},\bar{q}_{2}$.
$\mathcal{R}_{B_{j}}\left(p_{1},p_{2},p_{3};p_{B}\right)$ denotes
the probability density for a given $q_{1}q_{2}q_{3}$ with momenta
$p_{1}$, $p_{2}$ and $p_{3}$ forming a baryon $B_{j}$ with momentum
$p_{B}$. $\mathcal{R}_{M_{j}}\left(p_{1},p_{2};p_{M}\right)$ denotes
the probability density for a given $q_{1}\bar{q}_{2}$ with momenta
$p_{1}$ and $p_{2}$ forming a meson $M_{j}$ with momentum $p_{M}$. 

The hadronization is a non-perturbative process and the combination
kernel functions $\mathcal{R}_{B_{j}}$ and $\mathcal{R}_{M_{j}}$
are hard to determine from the first principle calculation at the
moment. Inspired by the quark number scaling property for hadronic
$p_{T}$ spectra at LHC \citep{Song:2017gcz,Gou:2017foe,Song:2019sez},
we can take the equal-velocity combination of constituent quarks and
antiquarks as the main feature of hadron formation. In this case,
we have

\begin{equation}
\mathcal{R}_{B_{j}}\left(p_{1},p_{2},p_{3};p_{B}\right)=\kappa_{B_{j}}\prod_{i=1}^{3}\delta\left(p_{i}-x_{i}p_{B}\right),\label{eq:RBj}
\end{equation}

\begin{equation}
\mathcal{R}_{M_{j}}\left(p_{1},p_{2};p_{M}\right)=\kappa_{M_{j}}\prod_{i=1}^{2}\delta\left(p_{i}-x_{i}p_{M}\right),\label{eq:RMj}
\end{equation}
Momentum fraction $x_{i}$ in baryon formula is $x_{i}=m_{i}/(m_{1}+m_{2}+m_{3})$
and that in meson formula is $x_{i}=m_{i}/(m_{1}+m_{2})$. $m_{i}$
is the constituent mass for quark of flavor $i$, and we take $m_{s}=$0.5
GeV, $m_{u}=m_{d}=$ 0.3 GeV. $\kappa_{B_{j}}$ and $\kappa_{M_{j}}$
are coefficients and independent of the momentum.

For the joint momentum distribution of quarks and antiquarks, we take
the factorization approximation, 

\begin{equation}
f_{q_{1}q_{2}q_{3}}\left(p_{1},p_{2},p_{3}\right)=f_{q_{1}}\left(p_{1}\right)f_{q_{2}}\left(p_{2}\right)f_{q_{3}}\left(p_{3}\right),\label{eq:fq1fq2fq3}
\end{equation}

\begin{equation}
f_{q_{1}\bar{q}_{2}}\left(p_{1},p_{2}\right)=f_{q_{1}}\left(p_{1}\right)f_{\bar{q}_{2}}\left(p_{2}\right).\label{eq:fq1fqbar2}
\end{equation}
Substituting Eqs.~(\ref{eq:RBj})-(\ref{eq:fq1fqbar2}) into Eqs.~(\ref{eq:fbi})
and (\ref{eq:fmi}), we obtain

\begin{equation}
f_{B_{j}}\left(p_{B}\right)=\kappa_{B_{j}}f_{q_{1}}\left(x_{1}p_{B}\right)f_{q_{2}}\left(x_{2}p_{B}\right)f_{q_{3}}\left(x_{3}p_{B}\right),\label{eq:fbi_indep}
\end{equation}

\begin{equation}
f_{M_{j}}\left(p_{M}\right)=\kappa_{M_{j}}f_{q_{1}}\left(x_{1}p_{M}\right)f_{\bar{q}_{2}}\left(x_{2}p_{M}\right).\label{eq:fmi_indep}
\end{equation}
We see that the momentum spectrum of hadron is simply the product
of those of quarks at hadronization. This simple form yields some
interesting flavor correlation among momentum distribution of different
hadrons, e.g, those at $p_{T}$ spectra \citep{Song:2017gcz,Song:2019sez}
and elliptic flows of different hadrons \citep{Song:2021ygg}. 

Coefficient $\kappa_{B_{j}}$ and $\kappa_{M_{j}}$ are independent
of momentum but dependent on quark numbers. To determine them and
clarify their physical meaning and importance, we write Eqs.~(\ref{eq:fbi_indep})
and~(\ref{eq:fmi_indep}) as 
\begin{equation}
f_{B_{j}}\left(p_{B}\right)=N_{q_{1}q_{2}q_{3}}\kappa_{B_{j}}f_{q_{1}}^{\left(n\right)}\left(x_{1}p_{B}\right)f_{q_{2}}^{\left(n\right)}\left(x_{2}p_{B}\right)f_{q_{3}}^{\left(n\right)}\left(x_{3}p_{B}\right),\label{eq:fb_nk}
\end{equation}

\begin{equation}
f_{M_{j}}\left(p_{M}\right)=N_{q_{1}\bar{q}_{2}}\kappa_{M_{j}}f_{q_{1}}^{\left(n\right)}\left(x_{1}p_{M}\right)f_{\bar{q}_{2}}^{\left(n\right)}\left(x_{2}p_{M}\right),\label{eq:fm_nk}
\end{equation}
by introducing the normalized quark distribution $f_{q_{i}}^{(n)}(p)=N_{q_{i}}f_{q_{i}}(p)$.
At the same time, we write hadron distribution as 
\begin{equation}
f_{B_{j}}\left(p_{B}\right)=N_{B_{j}}f_{B_{j}}^{\left(n\right)}\left(p_{B}\right),\label{eq:fbj_n}
\end{equation}

\begin{equation}
f_{M_{j}}\left(p_{M}\right)=N_{M_{j}}f_{M_{j}}^{\left(n\right)}\left(p_{M}\right).\label{eq:fmi_n}
\end{equation}
Comparing Eqs.~(\ref{eq:fb_nk}-\ref{eq:fm_nk}) and (\ref{eq:fbj_n}-\ref{eq:fmi_n}),
we obtain 
\begin{equation}
N_{B_{j}}=N_{q_{1}}N_{q_{2}}N_{q_{3}}\frac{\kappa_{B_{j}}}{A_{B_{j}}},\label{eq:NBj_p1}
\end{equation}

\begin{equation}
N_{M_{j}}=N_{q_{1}}N_{\bar{q}_{2}}\frac{\kappa_{M_{j}}}{A_{M_{j}}},\label{eq:NMj_p1}
\end{equation}
where 
\begin{equation}
A_{B_{j}}^{-1}=\int f_{q_{1}}^{\left(n\right)}\left(x_{1}p_{B}\right)f_{q_{2}}^{\left(n\right)}\left(x_{2}p_{B}\right)f_{q_{3}}^{\left(n\right)}\left(x_{3}p_{B}\right)dp_{B},\label{eq:A_Bj}
\end{equation}

\begin{equation}
A_{M_{j}}^{-1}=\int f_{q_{1}}^{\left(n\right)}\left(x_{1}p_{M}\right)f_{\bar{q}_{2}}^{\left(n\right)}\left(x_{2}p_{M}\right)dp_{M}.\label{eq:A_Mj}
\end{equation}
Clearly, $\kappa_{B_{j}}/A_{B_{j}}$ in Eq.~(\ref{eq:NBj_p1}) serves
as the momentum-integrated probability of $q_{1}q_{2}q_{3}$ forming
a baryon $B_{j}$. $\kappa_{M_{j}}/A_{M_{j}}$ in Eq.~(\ref{eq:NMj_p1})
denotes that of $q_{1}\bar{q}_{2}$ forming a meson $M_{j}$. Supposing
that quark system doubles in size, i.e., $N_{q_{i}}$ doubles, then
after hadronization $N_{h}$ should also double. However $N_{q_{1}}N_{q_{2}}N_{q_{3}}$
in Eq.~(\ref{eq:NBj_p1}) increases 8 times and $N_{q_{1}}N_{\bar{q}_{2}}$
in Eq.~(\ref{eq:NMj_p1}) increases 4 times. Therefore, probability
$\kappa_{B_{j}}/A_{B_{j}}$ in Eq.~(\ref{eq:NBj_p1}) and $\kappa_{M_{j}}/A_{M_{j}}$
in Eq.~(\ref{eq:NMj_p1}) should also play the role of the re-normalization
to guarantee the unitarity of the hadronization. Following this argument,
we can parameterize them as 

\begin{equation}
\kappa_{B_{j}}/A_{B_{j}}\equiv P_{q_{1}q_{2}q_{3}\shortrightarrow B_{j}}=C_{B_{j}}N_{iter}\frac{\overline{N}_{B}}{N_{qqq}},\label{eq:kappa_Bj_para}
\end{equation}

\begin{equation}
\kappa_{M_{j}}/A_{M_{j}}\equiv P_{q_{1}\bar{q}_{2}\shortrightarrow M_{j}}=C_{M_{j}}\frac{\overline{N}_{M}}{N_{q}N_{\bar{q}}}.\label{eq:kapa_Mj_para}
\end{equation}
 Here, $N_{qqq}=N_{q}(N_{q}-1)(N_{q}-2)\approx N_{q}^{3}$ with $N_{q}=N_{u}+N_{d}+N_{s}$
in heavy-ion collisions is the combination number of all $qqq$. $\overline{N}_{B}$
is the average number of all baryons. Then, $\overline{N}_{B}/N_{q}^{3}$
can denote the average probability of three quarks forming a baryon.
The factor $N_{iter}$ is the permutation number of $q_{1}q_{2}q_{3}$,
and equals to 1, 3, 6 for $q_{1}q_{2}q_{3}$ with identical flavors,
two identical flavors, and three different flavors, respectively.
$C_{B_{j}}$ is a refined parameter to account for the probability
of forming different spin state at quark combination. The meson formula
is similar. $\overline{N}_{M}/N_{q}N_{\bar{q}}$ denotes the average
probability of a $q\bar{q}$ pair forming a meson. $C_{M_{j}}$ account
for the probability of forming the meson with given spin state. 

In this paper, we only consider the production of ground state meson
$J^{P}=0^{-},1^{-}$ and baryon $J^{P}=\left(1/2\right)^{+},\left(3/2\right)^{+}$
in flavor SU(3) group. We introduce a parameter $R_{D/O}$ to denote
the relative probability of forming decuplet baryon to octet baryon
with the same quark flavor, and introduce a parameter $R_{V/P}$ to
denote the relative probability of vector meson to pseudo-scalar meson
with same quark flavor. Then $C_{B_{j}}$ and $C_{M_{j}}$ can be
written as 
\begin{equation}
C_{B_{j}}=\begin{cases}
\frac{1}{1+R_{D/O}} & for\ J^{P}=\left(1/2\right)^{+}\\
\frac{R_{D/O}}{1+R_{D/O}} & for\ J^{P}=\left(3/2\right)^{+}
\end{cases},\label{eq:Cbj}
\end{equation}
except $C_{\Lambda}=C_{\Sigma^{0}}=1/\left(2+R_{D/O}\right)$, \ $C_{\Sigma^{*0}}=R_{D/O}/\left(2+R_{D/O}\right),\ C_{\varDelta^{++}}=C_{\varDelta^{-}}=C_{\Omega^{-}}=1$,
and 
\begin{equation}
C_{M_{j}}=\begin{cases}
\frac{1}{1+R_{V/P}} & for\ J^{P}=0^{-}\\
\frac{R_{V/P}}{1+R_{V/P}} & for\ J^{P}=1^{-}
\end{cases}.
\end{equation}
Here, we take $R_{V/P}=0.55\pm0.05$ to reproduce the experimental
data of yield ratios $K^{\ast}/K$ and $\phi/K$ in high energy $pp$
and $p$Pb collisions \citep{Adam:2016bpr,Acharya:2019bli}. We take
$R_{D/O}=0.5\pm0.04$ by fitting the experimental data of yield ratios
$\Xi^{*}/\Xi$ and $\Sigma^{*}/\Lambda$ in high-energy $pp$ collisions
\citep{Adamova:2017elh}. 

For global production of baryons and mesons, i.e., $\overline{N}_{B}$
and $\overline{N}_{M}$, we have obtained their empirical solution
\citep{Song:2013isa},

\begin{align}
\overline{N}_{M} & =\frac{x}{2}\left[1-z\frac{\left(1+z\right)^{a}+\left(1-z\right)^{a}}{\left(1+z\right)^{a}-\left(1-z\right)^{a}}\right],\\
\overline{N}_{B} & =\frac{xz}{3}\frac{\left(1+z\right)^{a}}{\left(1+z\right)^{a}-\left(1-z\right)^{a}},\\
\overline{N}_{\bar{B}} & =\frac{xz}{3}\frac{\left(1-z\right)^{a}}{\left(1+z\right)^{a}-\left(1-z\right)^{a}},\label{eq:NBbar}
\end{align}
where $x=N_{q}+N_{\bar{q}}$ and $z=\left(N_{q}-N_{\bar{q}}\right)/x$.
$a=1+\left(\overline{N}_{M}/\overline{N}_{B}\right)_{z=0}/3$ characterizes
the production competition of baryon to meson at $z=0$ and is tuned
to be $a\approx4.86\pm0.1$ in relativistic heavy-ion collisions \citep{Shao:2017eok}.

When quark distributions $f_{q_{i}}\left(p\right)$ are given, momentum
distributions of hadrons $f_{h}\left(p\right)$ and their integrated
yields $N_{h}$ can be directly calculated using the model. In order
to compare with experimental data, the decay contribution of short-life
resonance should be also considered,

\begin{equation}
f_{h_{j}}^{\left(final\right)}\left(p\right)=f_{h_{i}}\left(p\right)+\sum_{i\neq j}\int dp'f_{h_{i}}\left(p'\right)D_{ij}\left(p',p\right).
\end{equation}
The decay function $D_{ij}\left(p',p\right)$ is determined by the
decay kinematics and decay branch ratios~\citep{olive2014chin}.

\section{TRANSVERSE MOMENTUM SPECTRUM OF STRANGE HADRONS\label{sec had_pt_spectra}}

In this section, we use the above EVC model to calculate $p_{T}$
spectra of strange hadrons at mid-rapidity in Au + Au collisions at
$\sqrt{s_{NN}}=$ 7.7, 11.5, 19.6, 27, 39 and 54.4 GeV, and make systematic
comparison with the experimental data \citep{adam2020strange,adamczyk2016probing,Huang:2021hbu,Huang:2022hbu,Muhammad:2019hbu}.
First, we reduce the formulas in the model to those in one-dimensional
$p_{T}$ space at mid-rapidity $y=0$. The momentum distribution function
$f(p)$ reduces to $f(p_{T})\equiv dN/dp_{T}$ and momentum integration
in Eqs. (\ref{eq:A_Bj}) and (\ref{eq:A_Mj}) reduce to $p_{T}$ integration. 

The model needs $p_{T}$ spectra of quarks and antiquarks at hadronization
$f_{q}(p_{T})$ as the input, which are difficult to obtain from first
principle calculations. Here, we take the following parametrization
form for the normalized quark $p_{T}$ spectrum

\begin{equation}
f_{q}^{\left(n\right)}\left(p_{T}\right)=\mathcal{N}_{q}\left(p_{T}+a_{q}\right)^{b_{q}}\left(1+\frac{\sqrt{p_{T}^{2}+m_{q}^{2}}-m_{q}}{n_{q}c_{q}}\right)^{-n_{q}},\label{eq:fqpt_par}
\end{equation}
where $\mathcal{N}_{q}$ is a normalized constant to assure $\int f_{q}^{\left(n\right)}\left(p_{T}\right)dp_{T}=1$.
Parameters $a_{q}$, $b_{q}$, $n_{q}$, $m_{q}$, $c_{q}$ control
the shape of the spectrum. We also need the numbers of quarks and
antiquarks $N_{q_{i}}$ so that $f_{q_{i}}(p_{T})=N_{q_{i}}f_{q_{i}}^{(n)}(p_{T})$
can be used to calculate $p_{T}$ spectra of hadrons in our model. 

\begin{figure*}
\centering{}\includegraphics[viewport=0bp 0bp 498bp 342bp,width=0.85\linewidth]{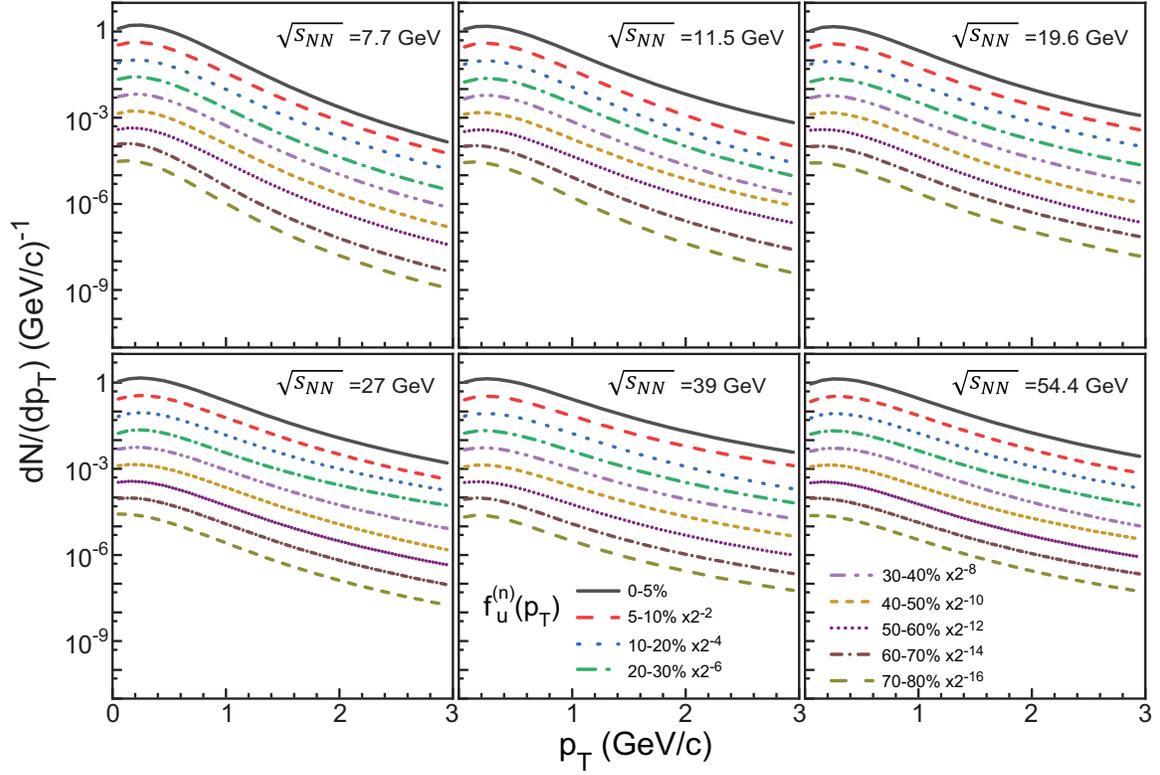}\caption{The $p_{T}$ spectra of $u$ quark in different centrality in Au +
Au collisions at $\sqrt{s_{NN}}$=7.7, 11.5, 19.6, 27, 39, 54.4 GeV.
\label{fig:fig1}}
\end{figure*}

\begin{figure*}
\centering{}\includegraphics[viewport=0bp 0bp 498bp 342bp,width=0.85\linewidth]{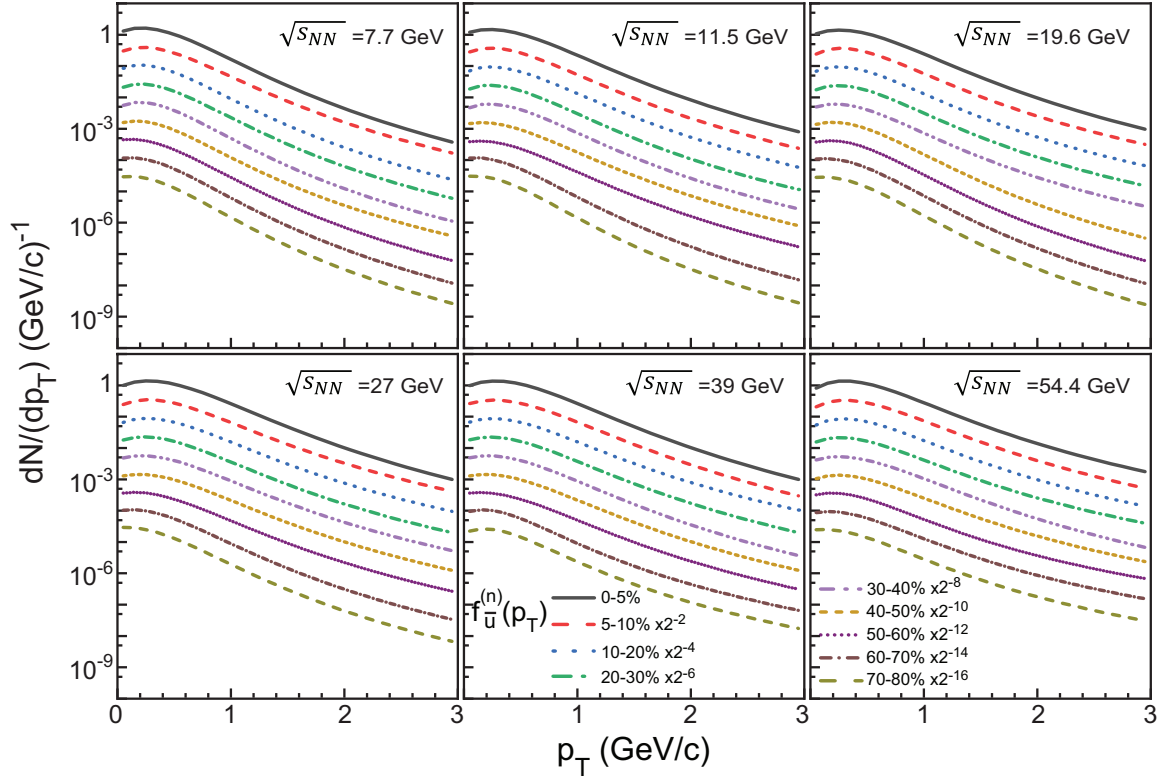}\caption{The same as \ref{fig:fig1} but for $\bar{u}$ . \label{fig:fig2}}
\end{figure*}

\begin{figure*}
\centering{}\includegraphics[viewport=0bp 0bp 498bp 342bp,width=0.85\linewidth]{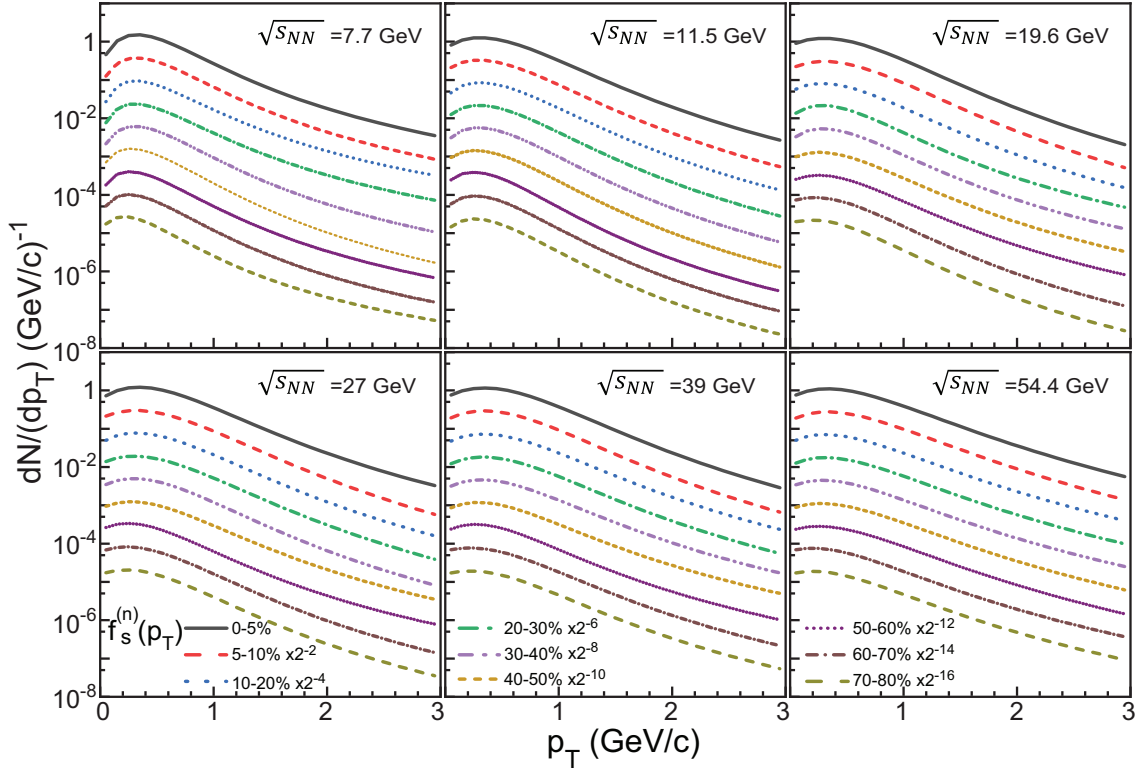}\caption{The same as \ref{fig:fig1} but for $s$. \label{fig:fig3}}
\end{figure*}

For hadron production at mid-rapidity at the studied collision energies,
we take the approximate isospin symmetry between up and down quarks,
i.e.,~$f_{u}(p_{T})=f_{d}(p_{T})$ and $f_{\bar{u}}(p_{T})=f_{\bar{d}}(p_{T})$.
We also assume the strangeness neutrality $f_{s}(p_{T})=f_{\bar{s}}(p_{T})$
according to our previous work \citep{Song:2020kak}. Finally, we
need three inputs $f_{u}(p_{T})$, $f_{\bar{u}}(p_{T})$ and $f_{s}(p_{T})$
to calculate $p_{T}$ spectra of light-flavor hadrons. They are fixed
by fitting experimental data of identified hadrons in our model. Specifically,
experimental data of $p_{T}$ spectrum of $\phi$ are used to fix
$f_{s}(p_{T})$, i.e., quark number $N_{s}$ and spectra parameters
$a_{s}$, $b_{s}$, $c_{s}$, $m_{s}$ and $n_{s}$. Because experimental
data for $p_{T}$ spectra of proton and antiproton are available only
at $p_{T}<2$ GeV/c \citep{STAR:2017sal}, they can only constrain
$p_{T}$ spectra of up quarks in a narrow range ($p_{T,u}\lesssim0.8$
GeV/c) and therefore we alternatively use experimental data for $p_{T}$
spectra of $\Lambda$ and $\bar{\Lambda}$ which cover a wider $p_{T}$
range to fix $f_{u}(p_{T})$ and $f_{\bar{u}}(p_{T})$, respectively.
Results for quark $p_{T}$ spectra at mid-rapidity in Au+Au collisions
in different centralities at $\sqrt{s_{NN}}=$ 7.7, 11.5, 19.6, 27,
39, 54.4 GeV are shown in Figs.~\ref{fig:fig1}-\ref{fig:fig3}.
Their properties are discussed latter in studying the baryon-to-meson
ratios in Sec. ~\ref{sec:BARYON TO MESON RATIOS } and nuclear modification
factors of hadrons in Sec.~\ref{sec:Rcp}. 

\begin{figure*}
\centering{}\includegraphics[width=0.85\linewidth]{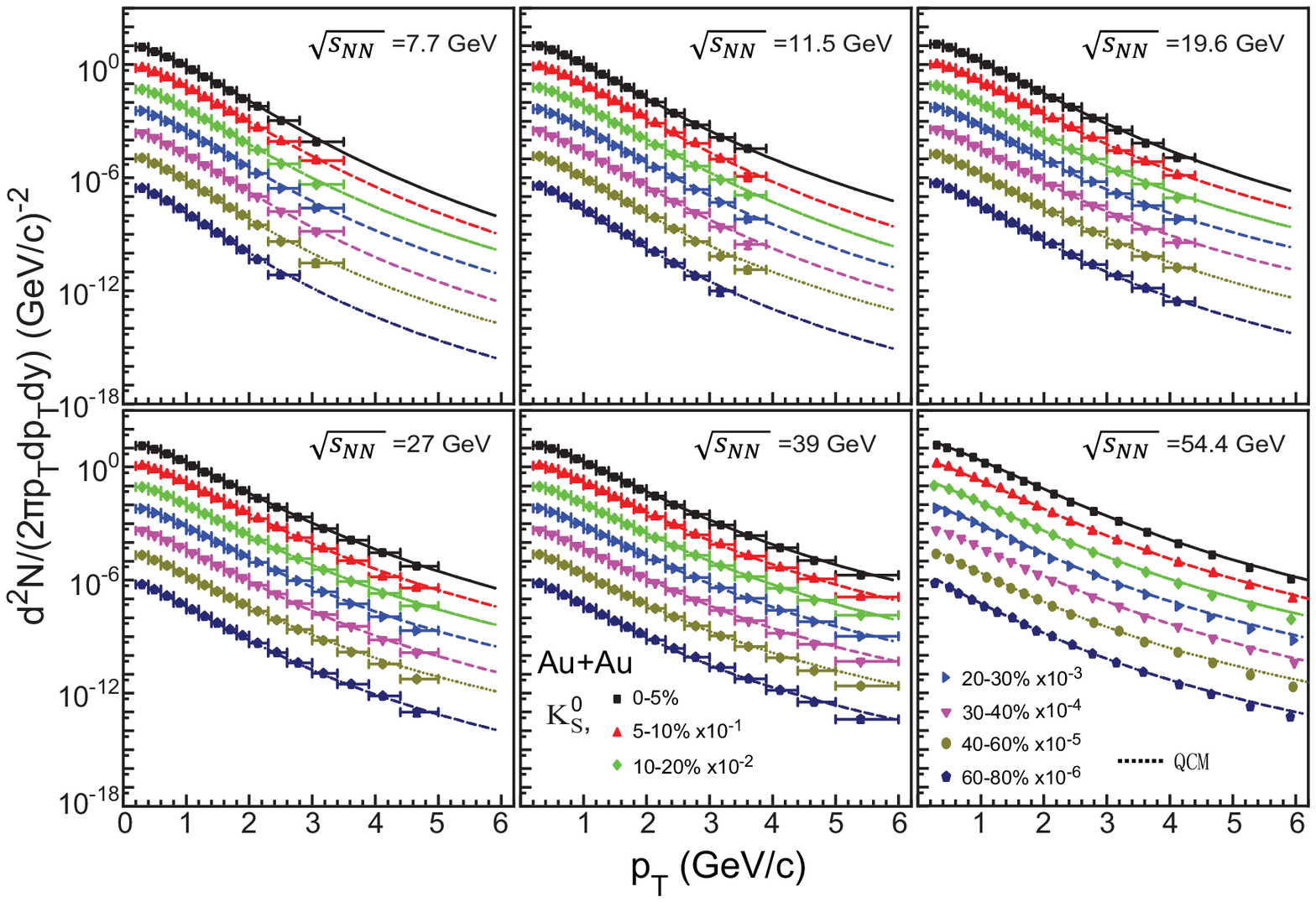}\caption{The $p_{T}$ spectra of $K_{S}^{0}$ at mid-rapidity ($|y|<0.5$)
from Au + Au collisions at $\sqrt{s_{NN}}$=7.7-54.4 GeV. Symbols
are the experimental data~\citep{adam2020strange,Huang:2021hbu,Muhammad:2019hbu}
and lines are the results of our model. Spectra of some hadrons are
scaled by factors 10 from central to peripheral collisions for clarity.
\label{fig:fig4}}
\end{figure*}

\begin{figure*}
\centering{}\includegraphics[width=0.85\textwidth]{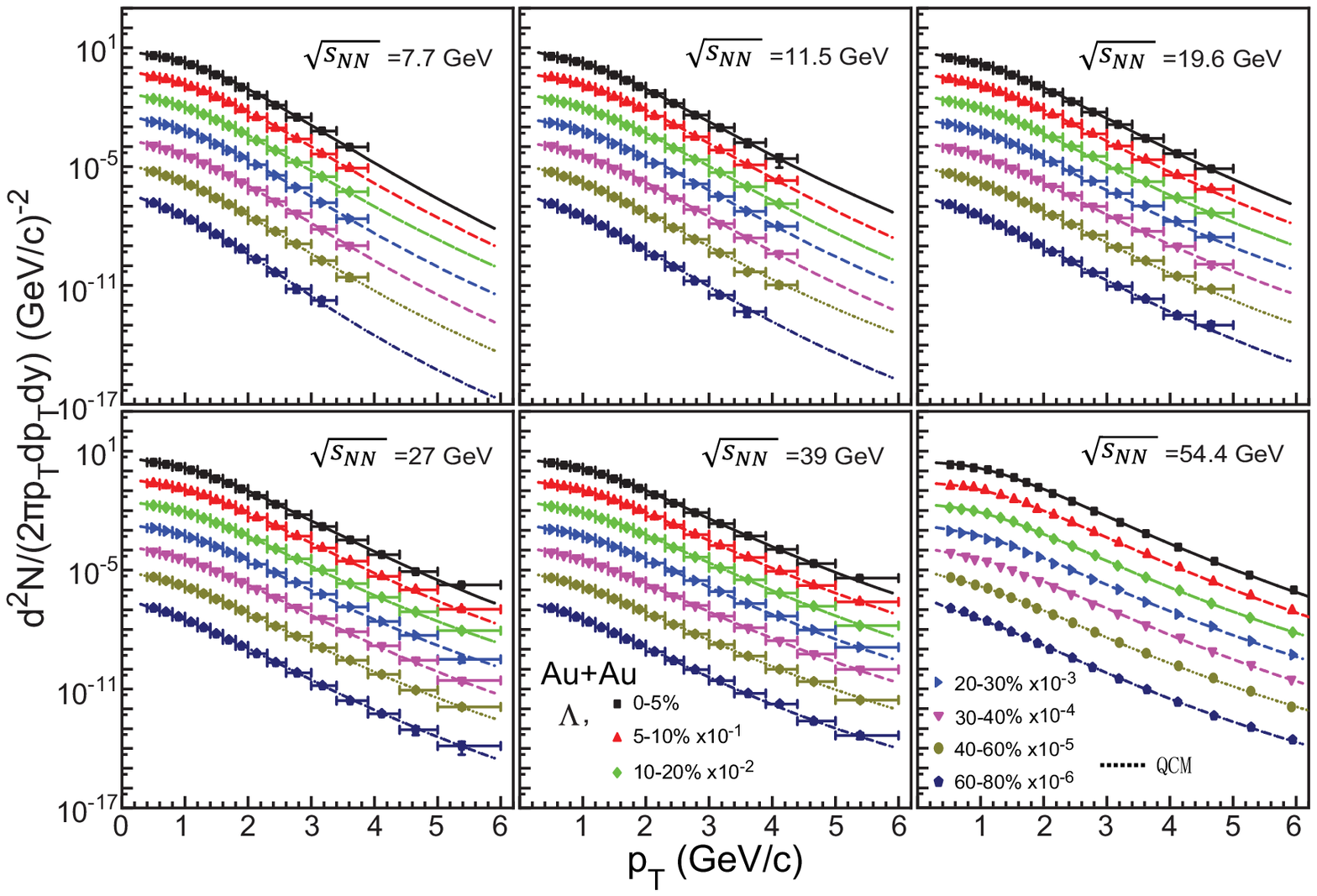}\caption{The same as Fig.~\ref{fig:fig4} but for $\Lambda$.\label{fig:fig5_Lam}}
\end{figure*}

\begin{figure*}
\centering{}\includegraphics[width=0.85\textwidth]{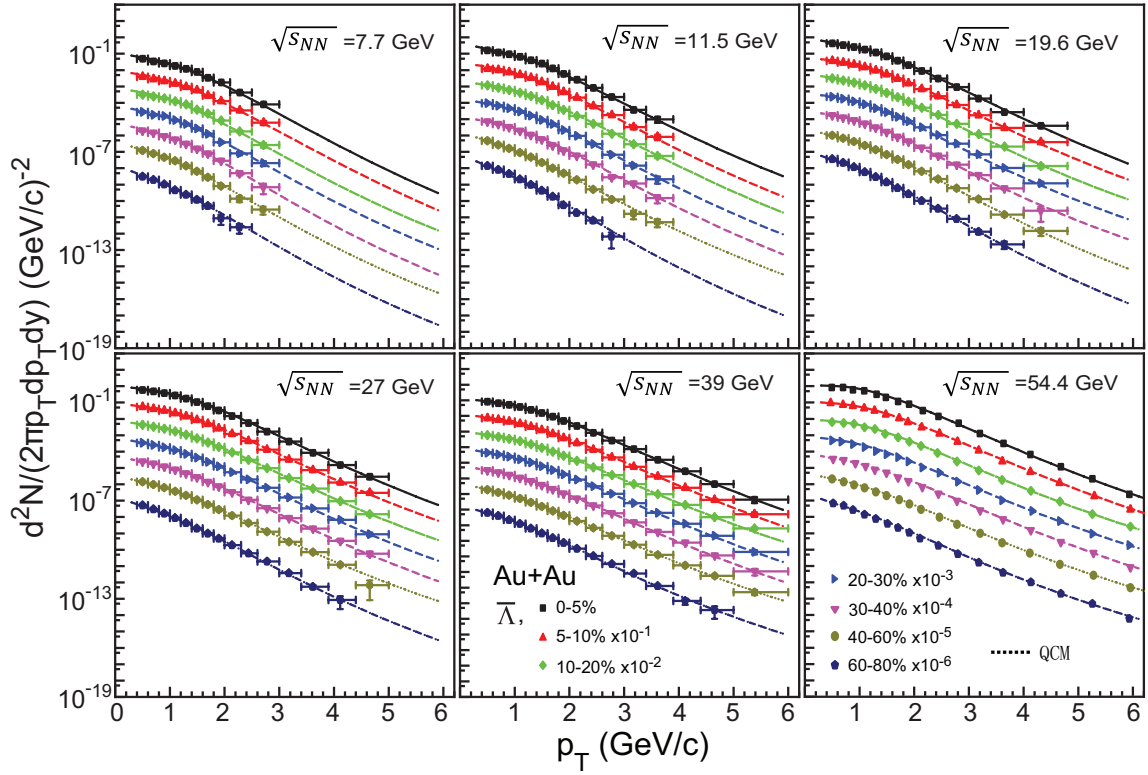}\caption{The same as Fig.~\ref{fig:fig4} but for $\bar{\Lambda}$.\label{fig:fig6_LamBar}}
\end{figure*}

\begin{figure*}
\centering{}\includegraphics[width=0.85\textwidth]{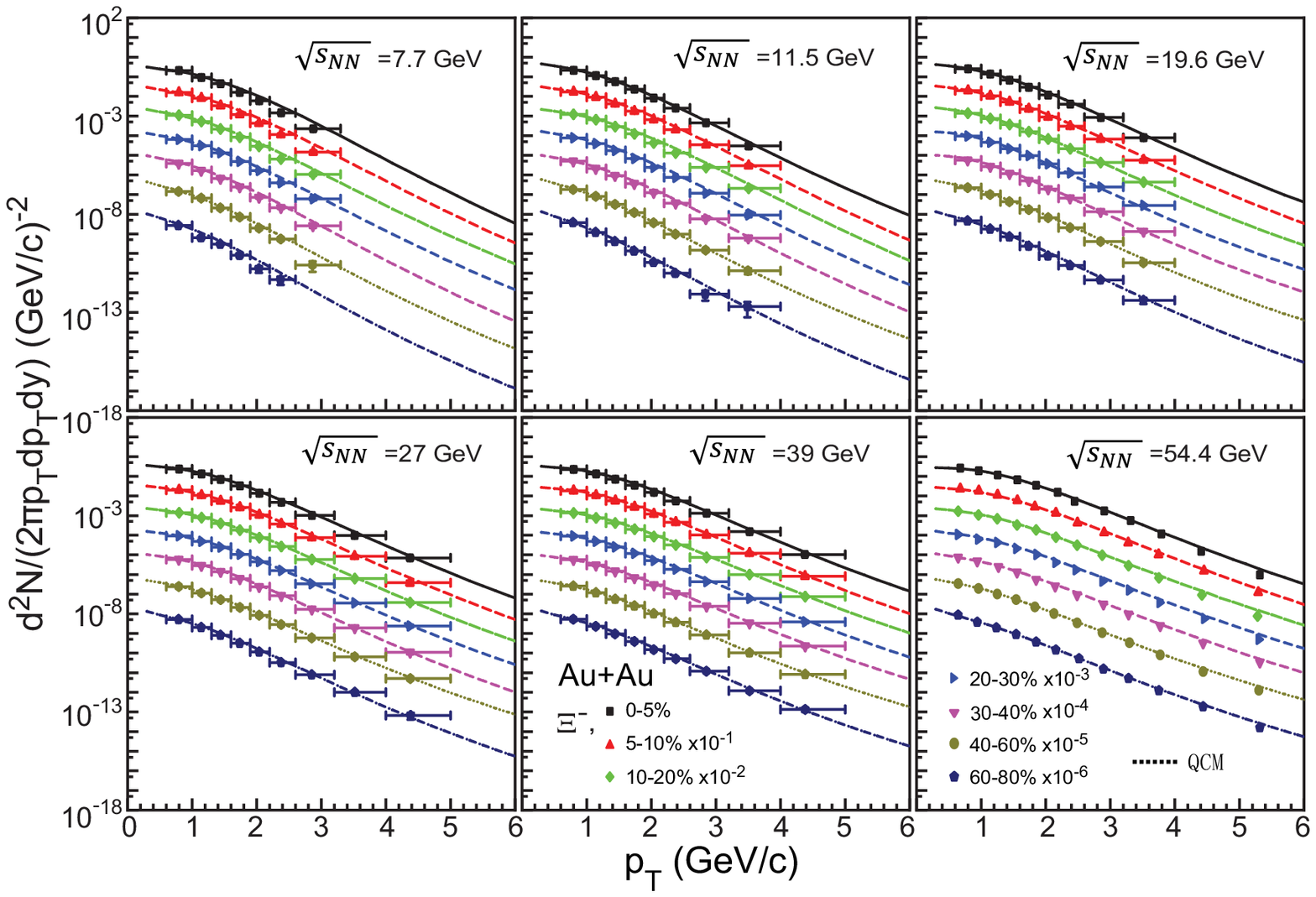}\caption{The same as Fig.~\ref{fig:fig4} but for $\Xi^{-}$.\label{fig:fig7}}
\end{figure*}

\begin{figure*}
\centering{}\includegraphics[width=0.85\textwidth]{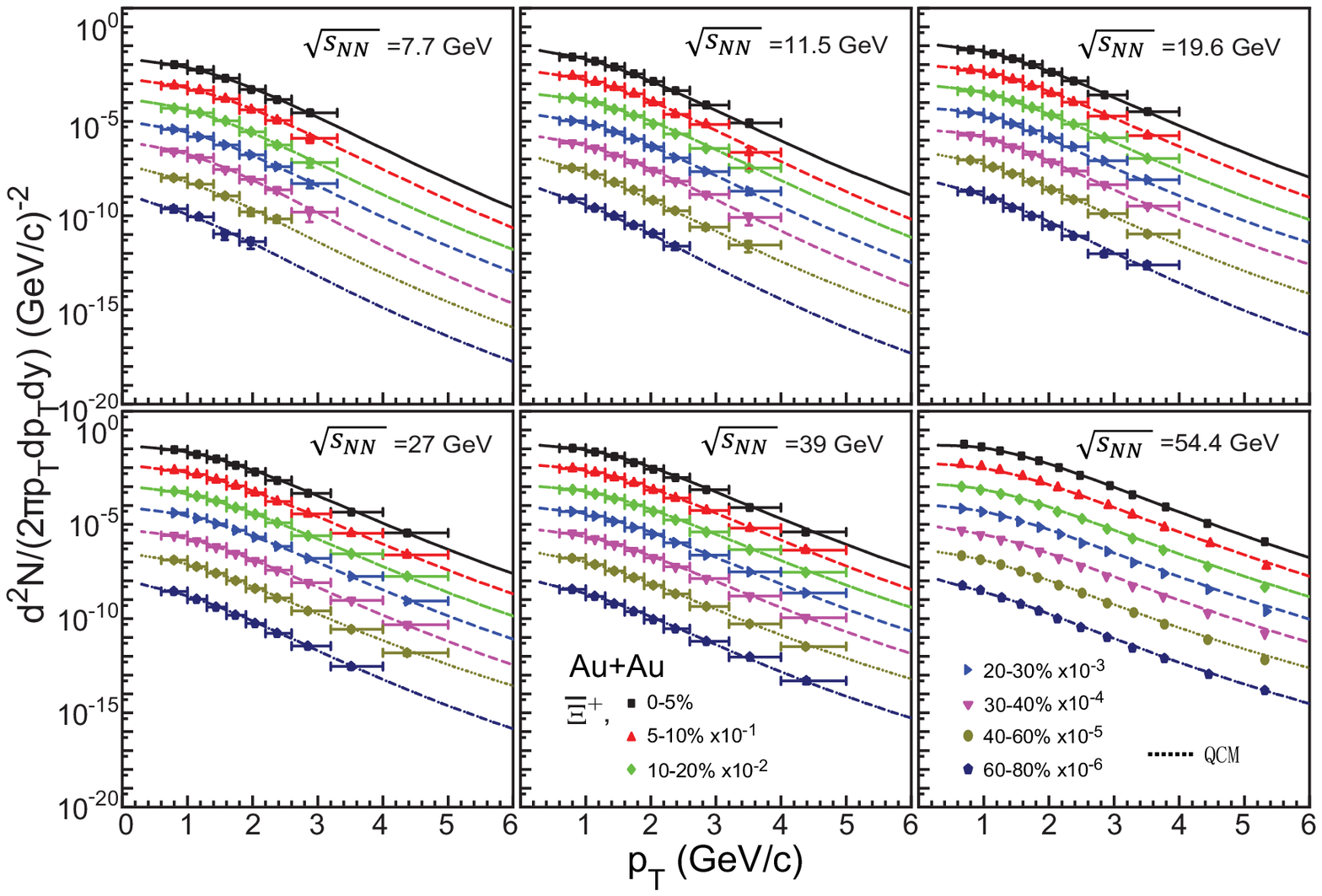}\caption{The same as Fig.~\ref{fig:fig4} but for $\bar{\Xi}^{+}$.\label{fig:Xibar_fpt}}
\end{figure*}

\begin{figure*}
\centering{}\includegraphics[width=0.85\textwidth]{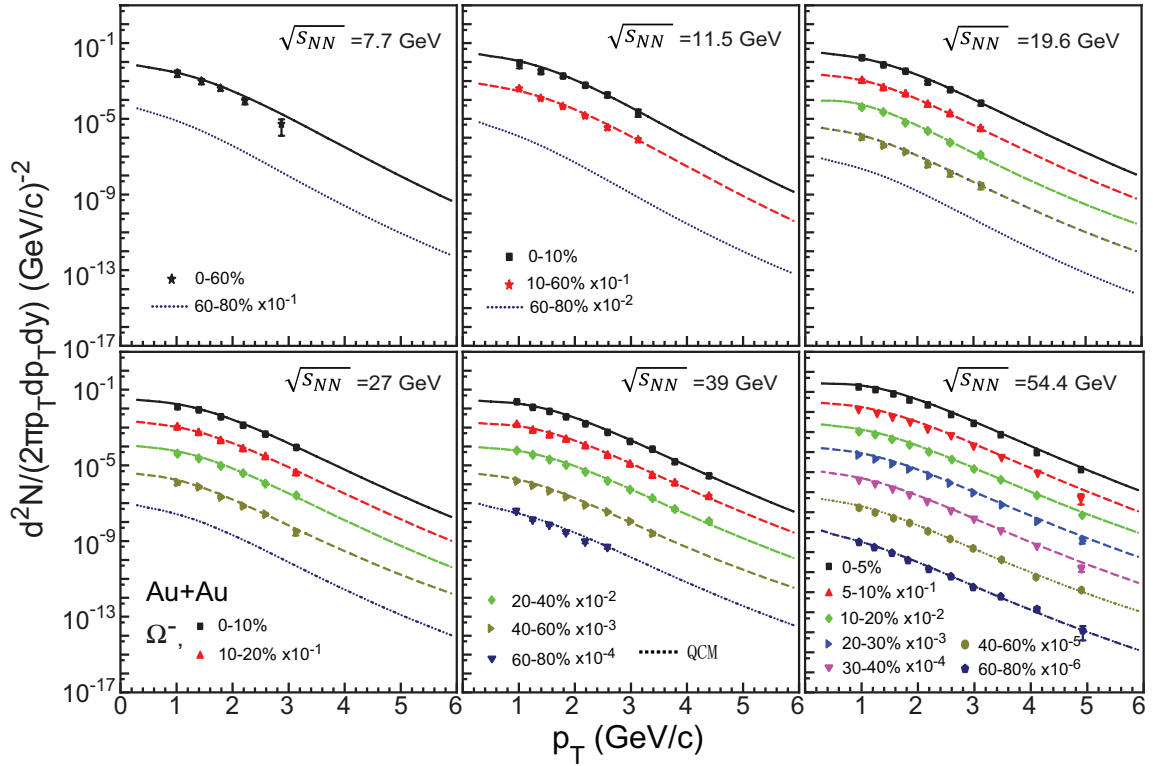}\caption{The same as Fig.~\ref{fig:fig4} but for $\Omega^{-}$. Experimental
data are from \citep{adamczyk2016probing,Huang:2021hbu}. \label{fig:fig9}}
\end{figure*}

Figs.~\ref{fig:fig4}-\ref{fig:fig11} show model results for $p_{T}$
spectra of different strange hadrons and their comparison with experimental
data \citep{adam2020strange,adamczyk2016probing,Huang:2021hbu,Muhammad:2019hbu}.
Besides three hadrons $\phi$, $\Lambda$, $\bar{\Lambda}$ which
are used to constrain quark $p_{T}$ spectra, results of other hadrons
$\Xi^{-}$, $\bar{\Xi}^{+}$, $\Omega^{-}$, $\bar{\Omega}^{+}$ as
well as $K_{S}^{0}$ \footnote{The $p_{T}$ spectrum of $K_{S}^{0}$ is calculated as the equal weight
mixing of that of $K^{0}$ and $\bar{K}^{0}$. In addition, because
kaon mass is smaller than the sum of constituent masses of down and
strange quarks used in this paper, the direct combination of down
and strange (anti-)quark is difficult to directly form the on-shell
kaon. Therefore, we modify the kaon formation in EVC mechanism as
follows: The combination of up/down and strange quark has a large
probability ($p_{1}$) to firstly form an intermediate resonance like
$K^{*}(892)$ and then decays into the on-shell kaon and pion; it
has a small probabilities ($p_{2}$) to directly from the on-shell
kaon. Here, we take $p_{1}=3/4$ and $p_{2}=1/4$ so that $p_{1}m_{K^{*}}+p_{2}m_{K}\approx0.8$
GeV approximately equals to $m_{u}+m_{s}$ and approximately satisfies
energy conservation. } are shown as theoretical prediction. In addition, we also calculate
$p_{T}$ spectra of proton and antiproton and find they are in good
agreement with the available data at $p_{T}<2$ GeV/c, which are not
shown here since we focus on strange hadrons in this paper. The systematic
comparison between model results of eight strange hadrons and their
experimental data can effectively test our model. 

Considering that the $p_{T}$ coverage and statistical uncertainties
of experiment data of these hadrons are different in different centralities
and/or at different collision energies, here in stead of the standard
$\chi^{2}/ndf$ evaluation, we use the relative deviation 

\begin{equation}
\overline{D}=\frac{1}{N_{data}}\sum_{i}^{N_{data}}\left|\frac{y_{i}^{(model)}-y_{i}^{(exp)}}{y_{i}^{(exp)}}\right|
\end{equation}
to quantify our model description. Here, $y_{i}^{(exp)}$ is the
central value of experimental data. Index $i$ runs over all datum
points of eight hadrons presented in Figs. \ref{fig:fig4}-\ref{fig:fig11}
in the given collision centrality and at the given collision energy.
Results of $\overline{D}$ are shown in Table.~\ref{tab1}.

We see that the relative deviation $\overline{D}$ in the studied
collision centralities and collision energy range is generally a few
percentages. $\overline{D}$ at $\sqrt{s_{NN}}=$19.6, 27, 39 and
54.4 GeV is about 0.02 in all collision centralities. $\overline{D}$
in semi-central and peripheral collisions at$\sqrt{s_{NN}}=$7.7 and
11.5 GeV is about 0.03-0.04, which is larger than, to a certain extent,
that in peripheral collisions at the two energies and those at higher
collision energies.

\begin{table*}[t]
\centering{}\caption{Relative deviation of theory and experiment for various
collision centralities and energy in Au+Au collisions at $\sqrt{s_{NN}}=$
7.7-54.4 GeV \citep{adam2020strange,adamczyk2016probing,Huang:2021hbu,Huang:2022hbu,Muhammad:2019hbu}.\label{tab1}}
\begin{tabular*}{11cm}{@{\extracolsep{\fill}}ccccccc}
\toprule 
Centrality & 7.7 & 11.5 & 19.6 & 27 & 39 & 54.4\tabularnewline
\midrule
\midrule 
0 \textminus{} 5\% & 0.0279 & 0.0170 & 0.0189 & 0.0166 & 0.0167 & 0.0214\tabularnewline
\midrule 
5 \textminus{} 10\% & 0.0442 & 0.0341 & 0.0265 & 0.0308 & 0.0240 & 0.0243\tabularnewline
\midrule 
10 \textminus{} 20\% & 0.0263 & 0.0226 & 0.0129 & 0.0121 & 0.0167 & 0.0226\tabularnewline
\midrule 
20 \textminus{} 30\% & 0.0265 & 0.0286 & 0.0174 & 0.0138 & 0.0120 & 0.0305\tabularnewline
\midrule 
30 \textminus{} 40\% & 0.0236 & 0.0285 & 0.0197 & 0.0134 & 0.0193 & 0.0248\tabularnewline
\midrule 
40 \textminus{} 60\% & 0.0412 & 0.0325 & 0.0159 & 0.0224 & 0.0181 & 0.0241\tabularnewline
\midrule 
60 \textminus{} 80\% & 0.0436 & 0.0411 & 0.0256 & 0.0205 & 0.0209 & 0.0228\tabularnewline
\bottomrule
\end{tabular*}
\end{table*}

The small values of $\overline{D}$ indicate that experimental data
of $p_{T}$ spectra of these eight hadrons in Au+Au collisions at
$\sqrt{s_{NN}}$=7.7-54.4 GeV can be self-consistently described by
our model. Such a global agreement therefore indicates the important
role of EVC mechanism at hadron production in these collisions. In
the following two sections, we further test our model by the baryon-to-meson
ratios and nuclear modification factors of identified hadrons, which
are more sensitive to hadron production mechanism in heavy-ion collisions. 

\begin{figure*}
\centering{}\includegraphics[width=0.85\textwidth]{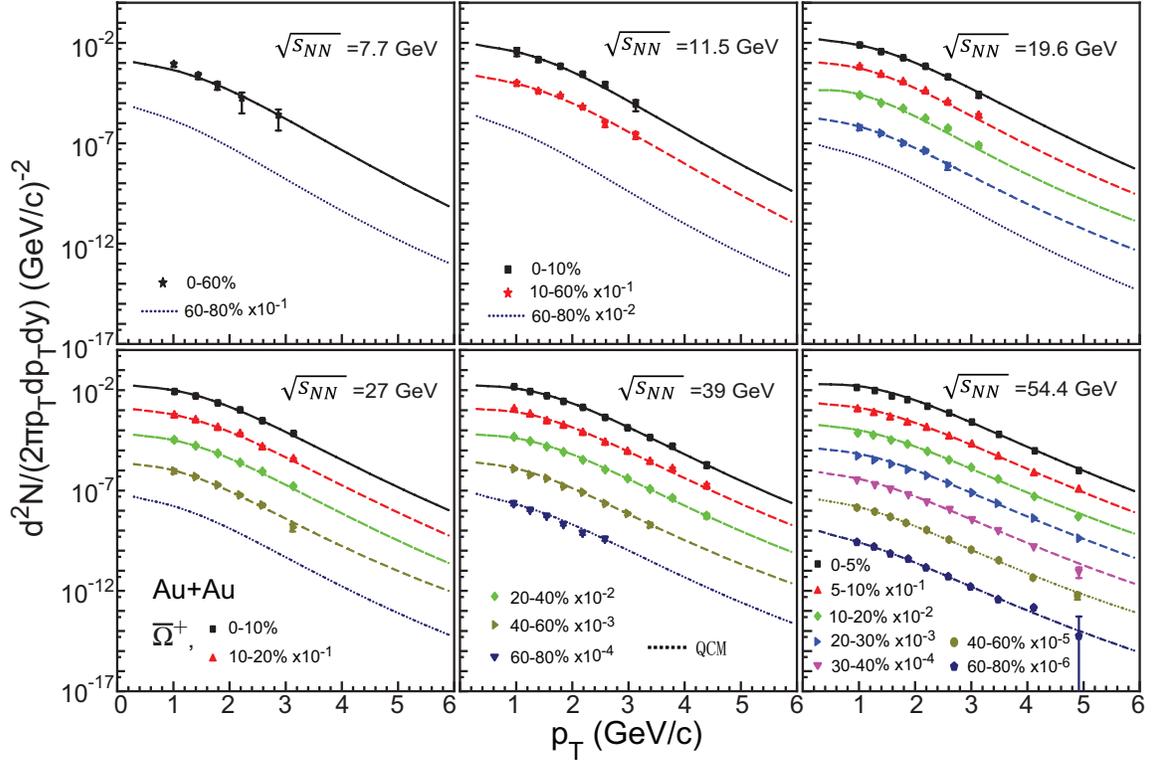}\caption{The same as Fig.~\ref{fig:fig9} but for $\bar{\Omega}^{+}$.}
\end{figure*}

\begin{figure*}
\centering{}\includegraphics[width=0.85\textwidth]{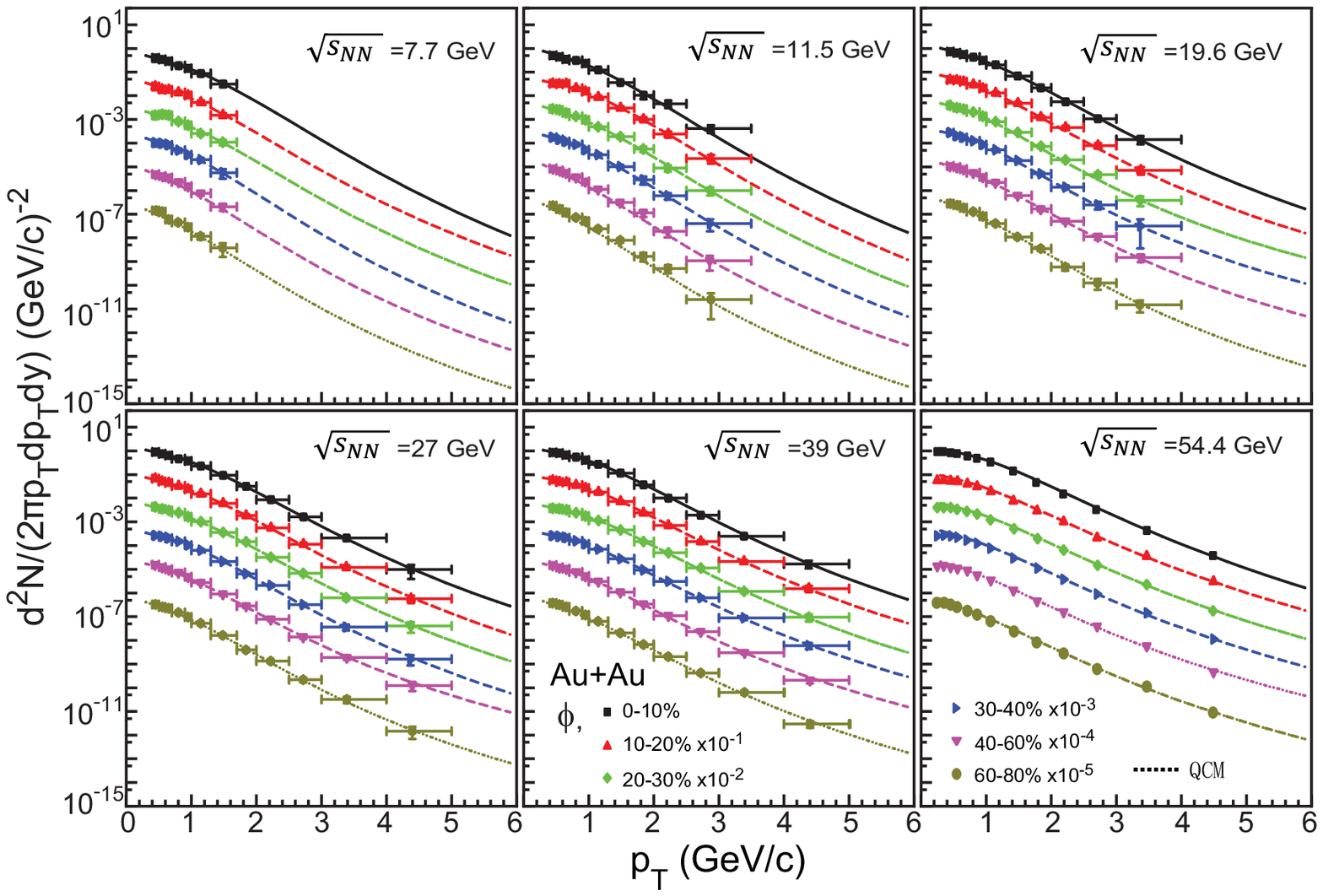}\caption{The same as Fig.~\ref{fig:fig9} but for $\phi$.\label{fig:fig11}}
\end{figure*}

\section{BARYON TO MESON RATIOS \label{sec:BARYON TO MESON RATIOS }}

Baryon-to-meson ratio is a sensitive physical quantity to check the
hadronization mechanism. RHIC experiments at the early years had shown
an enhancement of baryon-to-meson ratios at the intermediate $p_{T}$
in heavy-ion collisions. This enhancement is difficultly understood
in traditional fragmentation mechanism but can be naturally described
in quark (re-)combination mechanism. In this section, we study $\bar{\Lambda}/K_{S}^{0}$
and $\Omega/\phi$ ratios at/in different collision energies and/or
centralities. 

\begin{figure*}
\centering{}\includegraphics[width=1\linewidth]{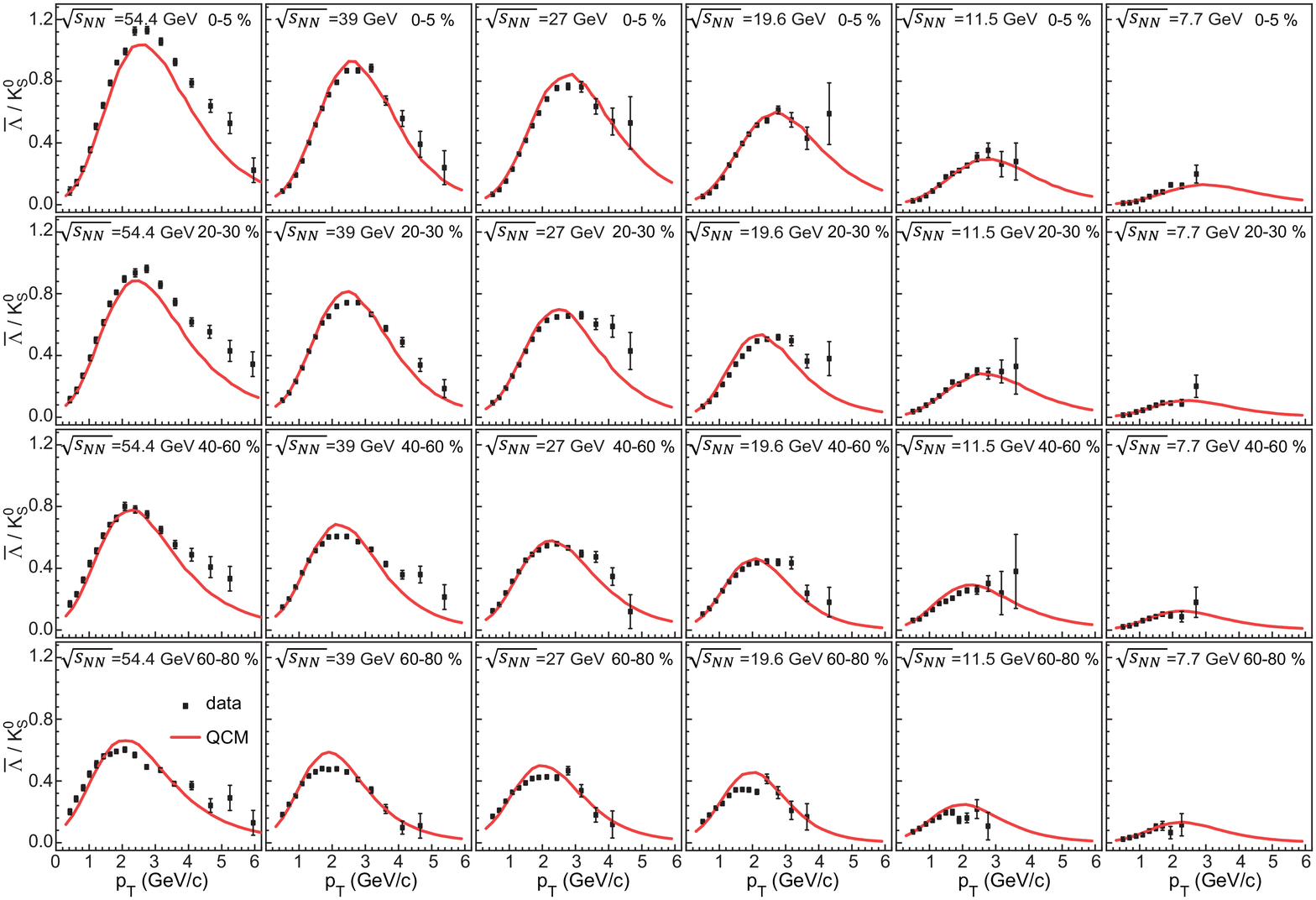}\caption{$\bar{\Lambda}$/$K_{S}^{0}$ ratio as a function of $p_{T}$ at mid-rapidity
($|y|<0.5$) for various collision centralities in Au + Au collisions
at $\sqrt{s_{NN}}$=7.7-54.4 GeV. Symbols are experimental data~\citep{adam2020strange,Muhammad:2019hbu}
and lines are results of our model.\label{fig:fig12}}
\end{figure*}

In Fig.~\ref{fig:fig12}, we show the results of $\bar{\Lambda}/K_{S}^{0}$
as the function of $p_{T}$ in Au+Au collisions with different centralities
at $\sqrt{s_{NN}}=7.7-54.4$ GeV and compare them with experimental
data~\citep{adam2020strange,Muhammad:2019hbu}. Experiment data of
$\bar{\Lambda}/K_{S}^{0}$ ratio exhibit an obvious dependence on
the collision energy and centrality. We see an obvious decrease of
the ratio with the decrease of collision energy. From central collisions
to peripheral collisions, we also see the decrease of the ratio. The
decrease magnitude of $\bar{\Lambda}/K_{S}^{0}$ with collision centrality
is generally slower than that with collision energy. For experimental
data of the central and semi-central collisions which have rich data
points and cover broad $p_{T}$ range, we always see a non-monotonic
$p_{T}$ dependence of the $\bar{\Lambda}/K_{S}^{0}$ ratio. The solid
lines are our model results. They are generally in good agreement
with experimental data. In the following text, we explain the underlying
physics for the $p_{T}$, collision energy and centrality dependence
of the ratio. 

Firstly, we explain the non-monotonic $p_{T}$ dependence of the $\bar{\Lambda}/K_{S}^{0}$
ratio. We examine the property of quark $p_{T}$ spectra shown in
Figs. \ref{fig:fig1}-\ref{fig:fig3} which is parameterized in Eq.~(\ref{eq:fqpt_par}).
In the range $p_{T,q}\lesssim1$ GeV/c (i.e., at low $p_{T}$ for
quarks), quark $p_{T}$ spectrum behaves approximately as 
\begin{equation}
dN_{q_{i}}/dp_{T}\propto p_{T}^{\alpha_{i}}\exp[-\sqrt{p_{T}^{2}+m_{i}^{2}}/T_{i}],\label{eq:fqpt_low_appr}
\end{equation}
where the exponent parameter $\alpha_{i}>0$ and slop parameter $T_{i}>0$.
Then $\bar{\Lambda}/K_{S}^{0}$ ratio in the low $p_{T}$ range ($p_{T}\approx2-3p_{T,q}\lesssim$
3 GeV/c)
\begin{align}
\frac{\Lambda}{K_{S}^{0}} & =\frac{\kappa_{\Lambda}f_{u}(x_{u}p_{T})f_{d}(x_{d}p_{T})f_{s}(x_{s}p_{T})}{\kappa_{K}f_{s}(x_{s}^{'}p_{T})f_{\bar{u}}(x_{u}^{'}p_{T})}\nonumber \\
 & =coe\times p_{T}^{\alpha_{u}}\label{eq:Lam_Ks0_low_pt}\\
 & \times\exp\left[-\sqrt{x_{u}^{2}p_{T}^{2}+m_{u}^{2}}\left(\frac{2}{T_{u}}+\frac{m_{s}/m_{u}}{T_{s}}\right)\right]\nonumber \\
 & \times\exp\left[+\sqrt{x_{u}^{'2}p_{T}^{2}+m_{u}^{2}}\left(\frac{1}{T_{\bar{u}}}+\frac{m_{s}/m_{u}}{T_{s}}\right)\right],\nonumber 
\end{align}
where we use $\alpha_{u}\approx\alpha_{\bar{u}}$ to simplify the
expression. Because the exponential terms change weakly with $p_{T}$,
the behavior of $\bar{\Lambda}/K_{S}^{0}$ ratio in the low $p_{T}$
range is therefore mainly determined by $p_{T}^{\alpha_{u}}$, which
is a rapidly increasing function. 

In the range $p_{T,q}\gtrsim1$ GeV/c, quark $p_{T}$ spectrum behaves
approximately as 
\begin{equation}
dN_{q_{i}}/dp_{T}\propto\left(1+\frac{p_{T}}{a_{i}}\right)^{-n_{i}},\label{eq:fqpt_high_appr}
\end{equation}
where the stretch parameter $a_{i}>0$ and and indices parameter $n_{i}>0$.
In the studied collision energies and centralities, $a$ is about
1-5 and $n_{i}$ is about 6-20. $\bar{\Lambda}/K_{S}^{0}$ ratio in
the range ($p_{T}\gtrsim$ 3 GeV/c)
\begin{align}
\frac{\Lambda}{K_{S}^{0}} & =\frac{\kappa_{\Lambda}f_{u}(x_{u}p_{T})f_{d}(x_{d}p_{T})f_{s}(x_{s}p_{T})}{\kappa_{K}f_{s}(x_{s}^{'}p_{T})f_{\bar{u}}(x_{u}^{'}p_{T})}\nonumber \\
 & \propto(a_{u}+x_{u}p_{T})^{-n_{u}}\label{eq:Lam_ks0_ratio_med_pt}\\
 & \times\left(1+\frac{\Delta x_{u}p_{T}}{a_{u}+x_{u}p_{T}}\right)^{n_{u}}\left(1+\frac{\Delta x_{s}p_{T}}{a_{s}+x_{s}p_{T}}\right)^{n_{s}},\nonumber 
\end{align}
where $\Delta x_{u}=x_{u}^{'}-x_{u}=0.1$ with $x_{u}=0.45$ and $\Delta x_{s}=x_{s}^{'}-x_{s}=0.17$
with $x_{s}=0.63$. Clearly, $(a_{u}+x_{u}p_{T})^{-n_{u}}$ is the
dominant term to drive $\Lambda/K_{S}^{0}$ ratio decrease with $p_{T}$
and terms $\left(1+\frac{\Delta x_{u}p_{T}}{a_{u}+x_{u}p_{T}}\right)^{n_{u}}$
and $\left(1+\frac{\Delta x_{s}p_{T}}{a_{s}+x_{s}p_{T}}\right)^{n_{s}}$
only weaken the influence of the first term to a certain. Combining
the effect of property of quark $p_{T}$ spectra in the low $p_{T}$
range in Eq.~ (\ref{eq:Lam_Ks0_low_pt}) and that in the $p_{T}$
range in Eq.~(\ref{eq:Lam_ks0_ratio_med_pt}), we now can understand
the increase of $\Lambda/K_{S}^{0}$ in the range $p_{T}\lesssim3$
GeV/c and subsequently its decrease in the range $p_{T}\gtrsim3$
GeV/c. 

\begin{figure}
\centering{}\includegraphics[viewport=0bp 0bp 440bp 210bp,width=1\linewidth]{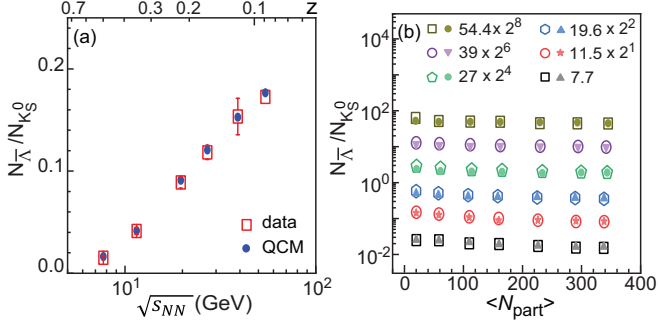}\caption{(a) The collision energy dependence of $N_{\bar{\Lambda}}/N_{K_{S}^{0}}$
ratios at mid-rapidity ($|y|<0.5$) in central Au + Au collisions
at $\sqrt{s_{NN}}$=7.7-54.4 GeV. The upper horizontal axis show $z$
of quarks; (b) $N_{\bar{\Lambda}}/N_{K_{S}^{0}}$ as functions of
$\left\langle N_{part}\right\rangle $ from Au+Au collisions at$\sqrt{s_{NN}}$=7.7-54.4
GeV. Open symbols are experimental data \citep{adam2020strange,Muhammad:2019hbu}
and solid symbols are model results .\label{fig:fig13}}
\end{figure}

Secondly, we explain that the energy dependence of the $\bar{\Lambda}/K_{S}^{0}$
ratio shown in Fig.~\ref{fig:fig12}, see the model calculation and
experimental data in a row. There are two main physical ingredients
that influence the $\bar{\Lambda}/K_{S}^{0}$ ratio. The first is
the relatively rapid increase of baryon chemical potential with the
decrease of collision energy. In our model, an asymmetry factor of
quark-antiquark number is defined as
\begin{equation}
z=\frac{N_{q}-N_{\bar{q}}}{N_{q}+N_{\bar{q}}},
\end{equation}
which closely relates to baryon chemical potential. $z$ causes the
production asymmetry between particles and antiparticles in our model
\citep{Song:2013isa}. At intermediate and low RHIC energies, $z$
is positive and is about $z\gtrsim0.1$. This will suppress the production
of anti-baryons. Therefore, the yield of $\bar{\Lambda}$ is largely
suppressed and this the main reason for the decrease of $\bar{\Lambda}/K_{S}^{0}$
with the decrease of collision energy shown in Fig.~\ref{fig:fig12}.
To illustrate it, we calculate the yield ratio of $\bar{\Lambda}$
to $K_{S}^{0}$ and after considering the resonance decays we obtain
\begin{equation}
\frac{N_{\bar{\Lambda}}}{N_{K_{S}^{0}}}=\frac{7.74}{(2+\lambda_{s})(1+0.12\lambda_{s})}R_{\bar{B}/M}(z),\label{eq:Nlamb_Nks0}
\end{equation}
where 
\begin{equation}
R_{\bar{B}/M}(z)=\frac{2z}{3(1+z)\left[(\frac{1+z}{1-z})^{a-1}-1\right]},
\end{equation}
with $a\approx4.86\pm0.1$ leading to $R_{\bar{B}/M}(z)\approx\frac{1}{11}-\frac{1}{12}$
in high energy collisions \citep{Song:2013isa,Shao:2017eok}. According
to the quark $p_{T}$ spectra in Figs.~\ref{fig:fig1}-\ref{fig:fig3},
we can calculate $z$ in at different collision energies and the results
of $\bar{\Lambda}/K_{S}^{0}$ yield ratio are shown in Fig.~\ref{fig:fig13}(a)
and compared with experimental data \citep{adam2020strange,Muhammad:2019hbu}.
We see that it decrease rapidly with the decrease of collision energy
(i.e., the increase of $z$). This is the main reason for the globally
rapid decrease of $\bar{\Lambda}/K_{S}^{0}$ ratio as the function
of $p_{T}$ shown in Fig.~\ref{fig:fig12}. Another reason that influences
the behavior of $\bar{\Lambda}/K_{S}^{0}$ ratio as the function of
$p_{T}$ is the shape change of quark $p_{T}$ spectra, which might
not be clearly seen from Figs.~\ref{fig:fig1}-\ref{fig:fig3}. Actually,
the extended $p_{T}$ range of thermal behavior of quark $p_{T}$
spectra in Eq.~\ref{eq:fqpt_low_appr} shrink with the decrease of
collision energy. This will also weaken the increasing trend of $\bar{\Lambda}/K_{S}^{0}$
ratio in the low $p_{T}$ range. 

Thirdly, we understand the centrality dependence of $\bar{\Lambda}/K_{S}^{0}$
ratio shown in Fig.~\ref{fig:fig12}, see the calculation results
and experimental data in a column. As shown in Fig.~\ref{fig:fig13}(b),
the $\bar{\Lambda}/K_{S}^{0}$ yield ratio changes weakly with collision
centrality at the studied collision energies. Therefore, the weak
change of $z$ contributes small centrality dependence to $\bar{\Lambda}/K_{S}^{0}$
ratio as the function of $p_{T}$. Actually, the main influence ingredient
comes from the change of quark $p_{T}$ spectra at different collision
centralities. From Figs.~\ref{fig:fig1}-\ref{fig:fig3}, we see
a clear shrink of thermal component for quark $p_{T}$ spectra in
peripheral collisions. This will cause the increase of $\bar{\Lambda}/K_{S}^{0}$
ratio in the low $p_{T}$ range stops at smaller $p_{T}$ in peripheral
collisions than that in central collisions. The maximum value of $\bar{\Lambda}/K_{S}^{0}$
ratio can reach in peripheral collisions also smaller than that in
central collisions. 

\begin{figure*}
\centering{}\includegraphics[width=1\textwidth]{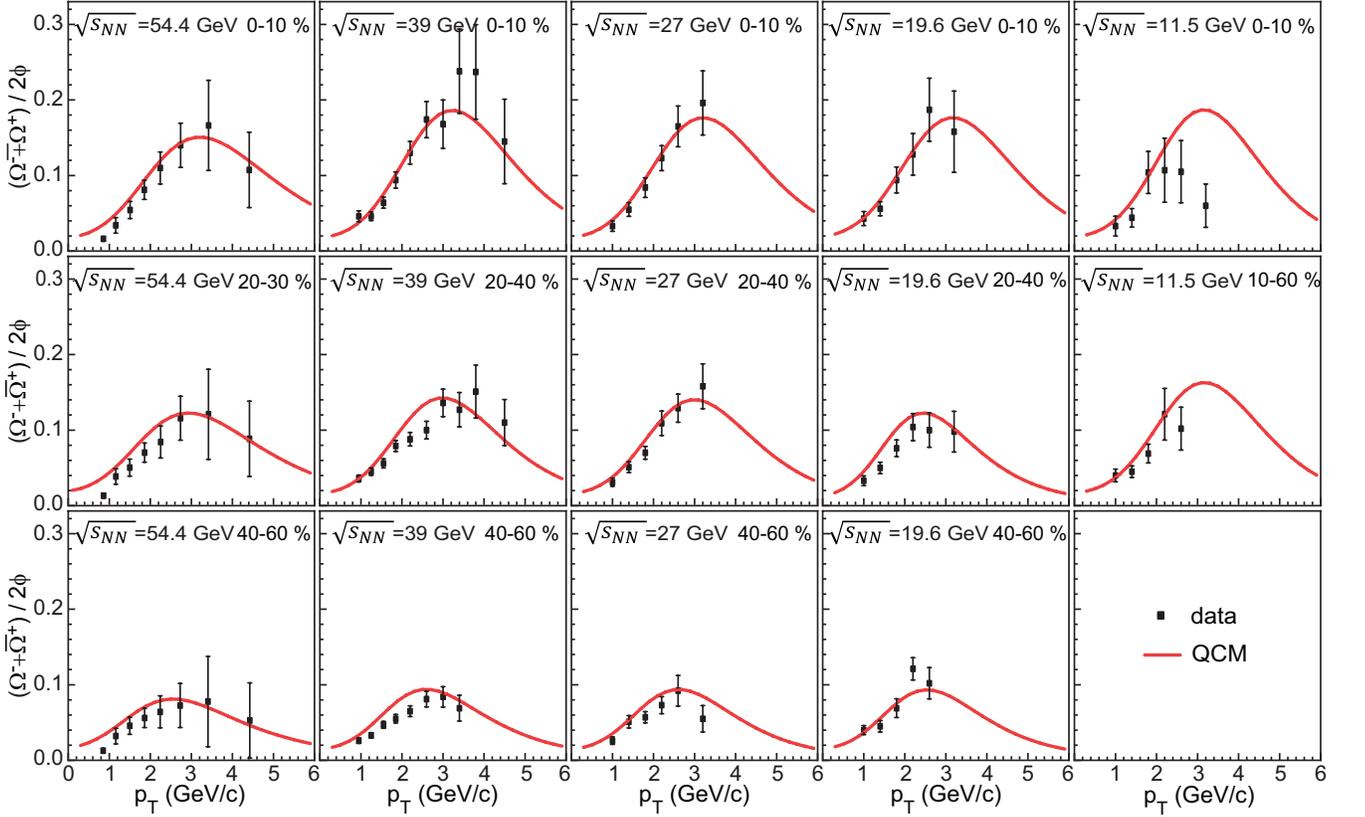}\caption{$\Omega/\phi$ ratio as a function of $p_{T}$ at mid-rapidity ($|y|<0.5$)
for various collision centralities in Au + Au collisions at $\sqrt{s_{NN}}$=11.5-54.4
GeV. Symbols are experimental data~\citep{adamczyk2016probing,Huang:2021hbu,Huang:2022hbu}
and lines are results of our model.\label{fig:fig14}}
\end{figure*}

Fig.~\ref{fig:fig14} show $\Omega/\phi$ ratio as the function of
$p_{T}$ in Au+Au collisions. Here, $\Omega$ denotes $\Omega^{-}+\bar{\Omega}^{+}$.
Symbols are experimental data ~\citep{adamczyk2016probing,Huang:2021hbu,Huang:2022hbu}
and lines are model results. We see that our model results are generally
in good agreement with the experimental data.

The underlying physics for the non-monotonic $p_{T}$ dependence of
$\Omega/\phi$ ratio in our model is quite similar with that discussed
in above $\bar{\Lambda}/K_{S}^{0}$ ratio. The difference in property
between quark $p_{T}$ spectra in the small $p_{T}$ range $p_{T}\lesssim$1
GeV/c and that in range $p_{T}\gtrsim$1 GeV/c leads to the $\Omega/\phi$
ratio firstly increases with $p_{T}$ and then decrease with $p_{T}$.
Because the production of $\Omega$ and $\phi$ involves only strange
(anti-)quarks, $p_{T}$ dependence of $\Omega/\phi$ ratio can be
seen more clear by the slope of the ratio
\begin{equation}
\left[\ln\frac{f_{\Omega}(p_{T})}{f_{\phi}(p_{T})}\right]^{'}=-\frac{1}{6}p_{T}\left[\ln f_{s}(\xi)\right]^{''},
\end{equation}
with $p_{T}/3<\xi<p_{T}/2$, which is obtained in our recent work
\citep{Li:2021nhq}. This equation means that the second derivative
of the logarithm of strange quark spectrum determine the increase
or decrease of the $\Omega/\phi$ ratio. In Fig.~\ref{fig:fig3}
with logarithmic vertical coordinate, we can intuitively see the sign
of $\left[\ln f_{s}(\xi)\right]^{''}$is negative in the range $p_{T,s}\lesssim1$
GeV/c and is positive as $p_{T,s}\gtrsim1$ GeV/c, which directly
causes the increase of $\Omega/\phi$ ratio in the range $p_{T}<2-3$
GeV/c and the decrease of the ratio as $p_{T}$ further increases. 

\begin{figure}[H]
\centering{}\includegraphics[viewport=0bp 0bp 430bp 210bp,width=1\linewidth]{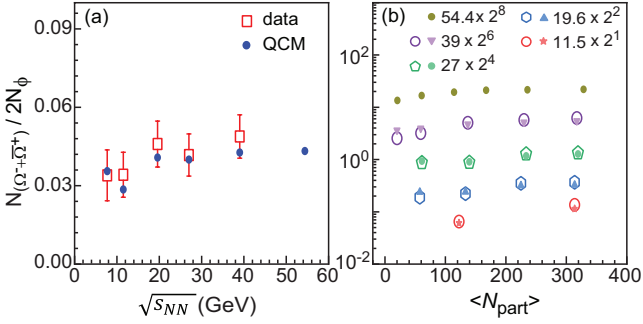}\caption{(a) The collision energy dependence of $N_{\Omega^{-}+\bar{\Omega}^{+}}/N_{\phi}$
ratios at mid-rapidity ($|y|<0.5$) in Au + Au collisions at $\sqrt{s_{NN}}$=7.7-54.4
GeV; (b) $N_{\Omega^{-}+\bar{\Omega}^{+}}/N_{\phi}$ as functions
of $\left\langle N_{part}\right\rangle $ from Au+Au collisions at
$\sqrt{s_{NN}}$=11.5-54.4 GeV. Open symbols are experimental data~\citep{adamczyk2016probing,Huang:2021hbu,Huang:2022hbu}
and solid symbols are model results.\label{fig:fig15}}
\end{figure}

The collision energy dependence of $\Omega/\phi$ ratio in Fig.~\ref{fig:fig14},
see results and data in a row, is not strong. This is mainly because
the $\Omega/\phi$ yield ratio, as shown in Fig.~\ref{fig:fig15}(a),
change weakly at the studied collision energies. The centrality dependence
of $\Omega/\phi$ ratio in Fig.~\ref{fig:fig14}, see results and
data in a column, is relatively obvious. This is mainly because of
the shape change of strange quark spectrum in different collision
centralities. From Fig.~\ref{fig:fig3}, we see a clear shrink of
thermal component for strange quark $p_{T}$ spectra in peripheral
collisions. This leads to the relatively weak increase of the $\Omega/\phi$
ratio in the low $p_{T}$ range and the relatively smaller $p_{T}$
at which $\Omega/\phi$ ratio begins to decrease in the peripheral
collisions. The centrality dependence of $\Omega/\phi$ yield ratio
is shown in Fig.~\ref{fig:fig15}(b). We see that $\Omega/\phi$
yield ratio in central collisions is larger than that in peripheral
collisions to a certain extent. This is because of the increase of
strange quarks fraction $\lambda_{s}$ in central collisions. This
will cause the global increase of $\Omega/\phi$ as the function of
$p_{T}$ in central collisions, in comparison with the $\Omega/\phi$
ratio in peripheral collisions. 

\section{NUCLEAR MODIFICATION FACTOR $R_{CP}$ \label{sec:Rcp}}

The nuclear modification factor ($R_{CP}$) of the final hadrons is
an important physical observable to quantify the difference between
hadron production in central collisions and that in peripheral collisions.
$R_{CP}$ is defined as 

\begin{equation}
R_{CP}(p_{T})=\frac{\left[\left(dN^{2}/2\pi p_{T}dp_{T}\right)/\left(N_{coll}\right)\right]_{central}}{\left[\left(dN^{2}/2\pi p_{T}dp_{T}\right)/\left(N_{coll}\right)\right]_{peripheral}}.\label{eq:Rcp_def}
\end{equation}
Here $N_{coll}$ is the number of binary nucleon-nucleon collisions
determined from Glauber model \citep{Miller:2007ri}. In general,
$R_{CP}$ of hadrons in relativistic heavy-ion collisions has a $p_{T}$
dependence. The property of $R_{CP}$ at high $p_{T}$ is driven by
jet quenching physics. In this paper, we focus on the property of
$R_{CP}$ of hadrons in the low and intermediate $p_{T}$ range, i.e.,
$p_{T}\lesssim4$ GeV/c for mesons and $p_{T}\lesssim$6 GeV/c for
baryons. The formation of hadrons in this range in our EVC mechanism
is by the combination of soft quarks and antiquarks with low transverse
momenta $p_{T,q}\lesssim2$ GeV/c.

Because $p_{T}$ spectra of hadrons in EVC mechanism exhibit relatively
simple relationship with those of quarks and antiquarks at hadronization,
$R_{CP}$ of hadrons can also exhibit some interesting properties
relating to quark flavor composition of hadron. Substituting Eq.~(\ref{eq:fbi_indep})
into Eq.~(\ref{eq:Rcp_def}), we obtain 
\begin{widetext}
\begin{align}
R_{CP,B_{j}} & (p_{T})=\frac{\frac{1}{N_{coll}^{(c)}}f_{B_{j}}^{(c)}(p_{T})}{\frac{1}{N_{coll}^{(p)}}f_{B_{j}}^{(p)}(p_{T})}\nonumber \\
 & =\frac{\frac{1}{N_{coll}^{(c)}}\kappa_{B_{j}}^{(c)}f_{q_{1}}^{(c)}(x_{q_{1}}p_{T})f_{q_{2}}^{(c)}(x_{q_{2}}p_{T})f_{q_{3}}^{(c)}(x_{q_{3}}p_{T})}{\frac{1}{N_{coll}^{(p)}}\kappa_{B_{j}}^{(p)}f_{q_{1}}^{(p)}(x_{q_{1}}p_{T})f_{q_{2}}^{(p)}(x_{q_{2}}p_{T})f_{q_{3}}^{(p)}(x_{q_{3}}p_{T})}\nonumber \\
 & =\left(\frac{N_{coll}^{(c)}}{N_{coll}^{(p)}}\right)^{2}\frac{\frac{N_{B}^{(c)}}{N_{q}^{(c)3}}A_{B_{j}}^{(c)}}{\frac{N_{B}^{(p)}}{N_{q}^{(p)3}}A_{B_{j}}^{(p)}}\frac{\left(\frac{1}{N_{coll}^{(c)}}\right)^{3}f_{q_{1}}^{(c)}(x_{q_{1}}p_{T})f_{q_{2}}^{(c)}(x_{q_{2}}p_{T})f_{q_{3}}^{(c)}(x_{q_{3}}p_{T})}{\left(\frac{1}{N_{coll}^{(p)}}\right)^{3}f_{q_{1}}^{(p)}(x_{q_{1}}p_{T})f_{q_{2}}^{(p)}(x_{q_{2}}p_{T})f_{q_{3}}^{(p)}(x_{q_{3}}p_{T})}\nonumber \\
 & =\left(\frac{N_{coll}^{(c)}}{N_{coll}^{(p)}}\right)^{2}\frac{\frac{N_{B}^{(c)}}{N_{q}^{(c)3}}A_{B_{j}}^{(c)}}{\frac{N_{B}^{(p)}}{N_{q}^{(p)3}}A_{B_{j}}^{(p)}}R_{CP,q_{1}}(x_{q_{1}}p_{T})R_{CP,q_{2}}(x_{q_{2}}p_{T})R_{CP,q_{3}}(x_{q_{3}}p_{T}).
\end{align}
\end{widetext}

\noindent In the third line, we have used Eq.~(\ref{eq:kappa_Bj_para}).
In the last line, we extend Eq.~(\ref{eq:Rcp_def}) to quarks at
hadronization. By rewriting
\begin{equation}
\frac{N_{B}}{N_{q}}=\frac{2}{1+z}\frac{z}{3}\frac{(1+z)^{a}}{(1+z)^{a}-(1-z)^{a}}=g_{B}(z),
\end{equation}
we finally obtain 
\begin{align}
 & R_{CP,B_{j}}(p_{T})\nonumber \\
 & =\frac{A_{B_{j}}^{(c)}g_{B}(z_{c})}{A_{B_{j}}^{(p)}g_{B}(z_{p})}\left(\frac{N_{coll}^{(c)}/N_{q}^{(c)}}{N_{coll}^{(p)}/N_{q}^{(p)}}\right)^{2}\label{eq:Rcp_decomp}\\
 & \times R_{CP,q_{1}}(x_{q_{1}}p_{T})R_{CP,q_{2}}(x_{q_{2}}p_{T})R_{CP,q_{3}}(x_{q_{3}}p_{T}).\nonumber 
\end{align}
We see that $R_{CP}$ of baryons directly relates to the product of
those of quarks. $g_{B}(z_{c})/g_{B}(z_{p})$ is only slightly smaller
than one because $z_{c}$ in central collisions is slightly greater
than $z_{p}$ in peripheral collisions. Coefficient $A_{B_{j}}^{(c)}/A_{B_{j}}^{(p)}$
is slightly smaller than one because quark $p_{T}$ spectra become
steeper to a certain extend when collision impact factor becomes large.
Coefficient $\left(\frac{N_{coll}^{(c)}/N_{q}^{(c)}}{N_{coll}^{(p)}/N_{q}^{(p)}}\right)^{2}$
is slightly greater than one because $N_{coll}\propto N_{part}^{4/3}$
in Glauber model \citep{Miller:2007ri} and $N_{q}\propto N_{part}$
in our model. Therefore, coefficients in right hand side of Eq.~(\ref{eq:Rcp_decomp})
is about one and the product of $R_{CP}$ of quarks dominates $R_{CP}$
of baryons. 

Applying Eq.~(\ref{eq:Rcp_decomp}) to ${\color{red}\Omega}$, we
have 
\begin{equation}
R_{CP,\Omega}(p_{T})=\frac{A_{\Omega}^{(c)}g_{B}(z_{c})}{A_{\Omega}^{(p)}g_{B}(z_{p})}\left(\frac{N_{coll}^{(c)}/N_{q}^{(c)}}{N_{coll}^{(p)}/N_{q}^{(p)}}\right)^{2}R_{CP,s}^{3}\left(\frac{p_{T}}{3}\right).\label{eq:RCP_Omega}
\end{equation}
A similar derivation for $\phi$ meson gives 
\begin{equation}
\begin{split}R_{CP,\phi}(p_{T})=\frac{A_{\phi}^{(c)}g_{M}(z_{c})}{A_{\phi}^{(p)}g_{M}(z_{p})}\left(\frac{N_{coll}^{(c)}/N_{q}^{(c)}}{N_{coll}^{(p)}/N_{q}^{(p)}}\right)R_{CP,s}^{2}\left(\frac{p_{T}}{2}\right),\end{split}
\label{eq:RCP_phi}
\end{equation}
where $g_{M}(z)=\dfrac{1}{1-z}\left[1-z\dfrac{\left(1+z\right)^{a}+\left(1-z\right)^{a}}{\left(1+z\right)^{a}-\left(1-z\right)^{a}}\right]$. 

\begin{figure*}
\centering{}\includegraphics[width=0.75\linewidth]{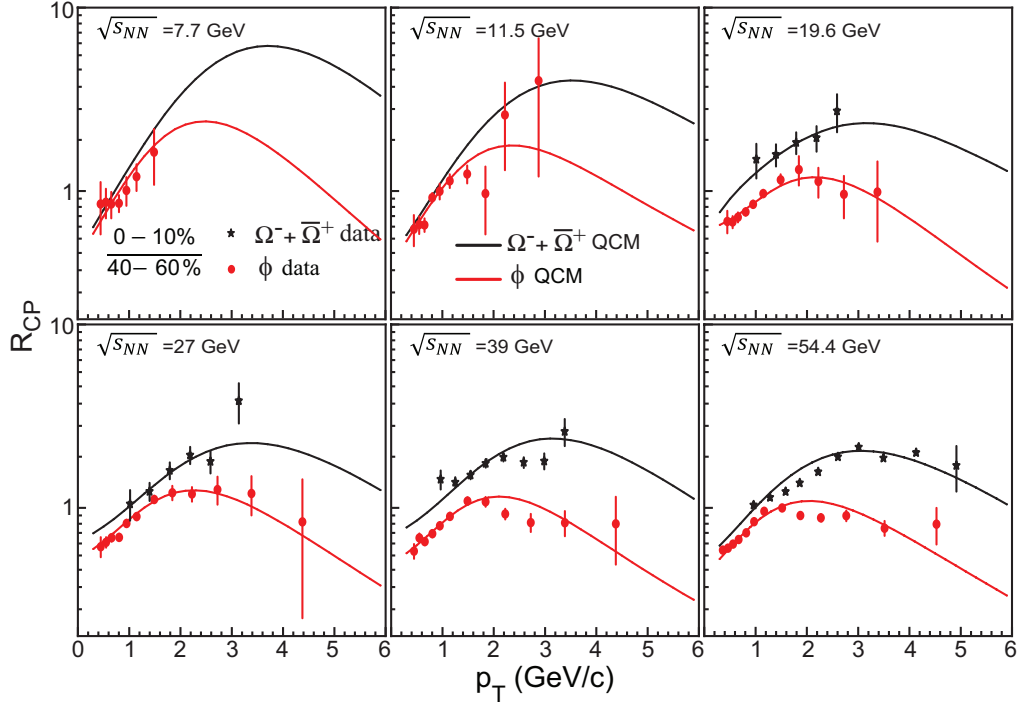}\caption{$R_{CP}$(0-10\%)/(40-60\%) of $\Omega^{-}+\bar{\Omega}^{+}$, $\phi$
, at mid-rapidity ($|y|<0.5$) in Au+Au collisions at $\sqrt{s_{NN}}$=7.7-54.4
GeV. Symbols are experimental data~\citep{adamczyk2016probing,Huang:2021hbu,Huang:2022hbu}
and lines are results of our model.\label{fig:fig16}}
\end{figure*}

Fig.~\ref{fig:fig16} show $R_{CP}$ of $\Omega$ and $\phi$ between
centrality $0-10\%$ and centrality $40-60\%$. Symbols are experimental
data~\citep{adamczyk2016probing,Huang:2021hbu,Huang:2022hbu} and
lines are model results which are directly calculated from the numerical
results of $p_{T}$ spectra of $\Omega$ and $\phi$ in Figs.~\ref{fig:fig9}-\ref{fig:fig11}.
We see that model results in Au+Au collisions at $\sqrt{s_{NN}}=$
19.6, 27, 39 and 54.4 GeV are in good agreement with experimental
data. 

\begin{figure}[H]
\centering{}\includegraphics[width=0.95\linewidth]{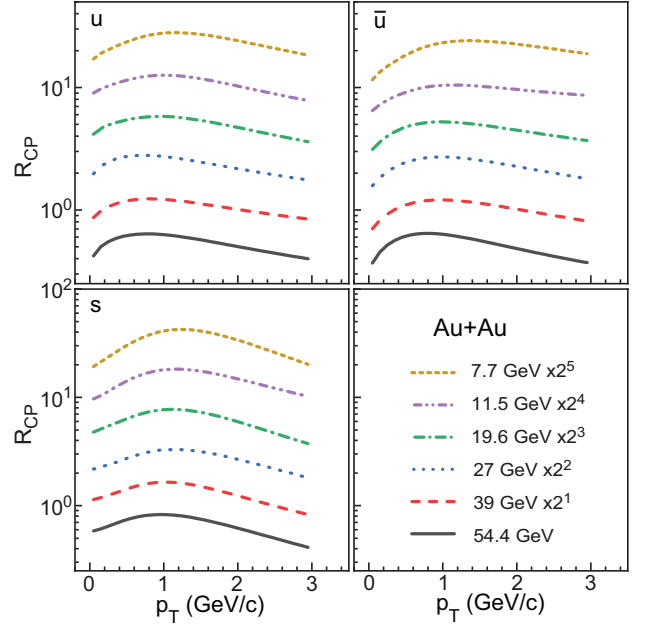}\caption{$R_{CP}$(0-5\%)/(40-60\%) of $u$, $\bar{u}$ and $R_{CP}$(0-10\%)/(40-60\%)
of $s$ at mid-rapidity ($|y|<0.5$) in Au+Au collisions at $\sqrt{s_{NN}}$=7.7-54.4
GeV.\label{fig:fig17}}
\end{figure}

Applying Eqs. (\ref{eq:RCP_Omega}) and (\ref{eq:RCP_phi}), we can
naturally explain the different $p_{T}$ dependence of experimental
data for $R_{CP}$ of $\Omega^{-}$ and $\phi$. As shown by Eqs.
(\ref{eq:RCP_Omega}) and (\ref{eq:RCP_phi}), $R_{CP,\Omega}(p_{T})$
and $R_{CP,\phi}(p_{T})$ in the EVC mechanism relate to the third
and second power of $R_{CP,s}(p_{T})$. Now, we examine the behavior
of strange quark $R_{CP,s}(p_{T})$, which can be calculated with
Eq. (\ref{eq:Rcp_def}) by strange quark $p_{T}$ spectra in Fig.~\ref{eq:kappa_Bj_para}
and the results are shown in Fig. \ref{fig:fig17}. We see that $R_{CP,s}(p_{T})$
increases with $p_{T}$ in the range $0<p_{T}\lesssim1$ GeV/c and
turns to decrease with $p_{T}$ in the range $p_{T}\gtrsim1$ GeV/c.
Because $p_{T,\phi}=2p_{T,s}$, $R_{CP,\phi}(p_{T})$ should increase
with $p_{T}$ in the range $p_{T}\lesssim2$ GeV/c and then turns
to decrease with $p_{T}$ as $p_{T}\gtrsim2$ GeV/c. Because $p_{T,\Omega}=3p_{T,s}$,
$R_{CP,\Omega}(p_{T})$ should increase with $p_{T}$ in the range
$p_{T}\lesssim3$ GeV/c and turns to decrease with $p_{T}$ as $p_{T}\gtrsim3$
GeV. Moreover, because $R_{CP,\Omega}(p_{T})$ relates to the third
power of $R_{CP,s}(p_{T})$ but $R_{CP,\phi}(p_{T})$ relates to the
square of $R_{CP,s}(p_{T})$, $R_{CP,\Omega}(p_{T})$ can not only
keep the increase trend in the larger $p_{T}$ range but also reach
higher magnitude, which are just seen in experimental data. 

\begin{figure}[H]
\centering{}\includegraphics[width=1\linewidth]{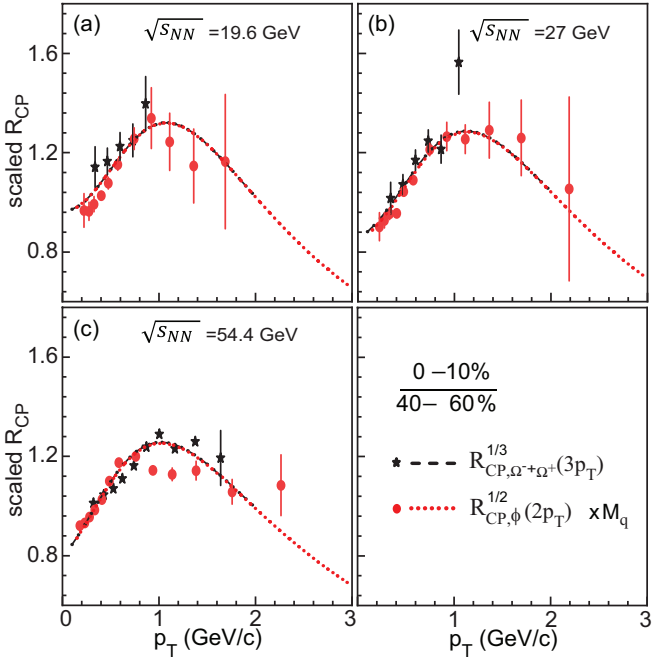}\caption{The scaling property for $R_{CP}$(0-10\%)/(40-60\%) of $\Omega^{-}+\bar{\Omega}^{+}$
and $\phi$ at mid-rapidity ($|y|<0.5$) in Au+Au collisions at $\sqrt{s_{NN}}$=7.7-54.4
GeV. Symbols are experimental data \citep{adamczyk2016probing,Huang:2021hbu,Huang:2022hbu}
and lines are results of our model. The coefficient $M_{q}$ is 1.16,
1.15 and 1.18, respectively. \label{fig:fig18}}
\end{figure}

In view of this good agreement and simple expressions for $\Omega^{-}$
and $\phi$ in Eqs.~(\ref{eq:RCP_Omega}) and (\ref{eq:RCP_phi}),
we can further build a correlation 
\begin{equation}
R_{CP,\Omega}^{1/3}(3p_{T})=M_{q}R_{CP,\phi}^{1/2}(2p_{T}),\label{eq:RCP_Omg_phi_Scale}
\end{equation}
where 
\begin{equation}
M_{q}=\dfrac{A_{\Omega}^{(c)\frac{1}{3}}A_{\phi}^{(p)\frac{1}{2}}}{A_{\Omega}^{(p)\frac{1}{3}}A_{\phi}^{(c)\frac{1}{2}}}\dfrac{g_{B}^{\frac{1}{3}}(z_{c})g_{M}^{\frac{1}{2}}(z_{p})}{g_{B}^{\frac{1}{3}}(z_{p})g_{M}^{\frac{1}{2}}(z_{c})}\left(\dfrac{N_{coll}^{(c)}/N_{q}^{(c)}}{N_{coll}^{(p)}/N_{q}^{(p)}}\right)^{\frac{1}{6}}
\end{equation}
 is close to one. Fig. \ref{fig:fig18} show the results of Eq.~(\ref{eq:RCP_Omg_phi_Scale})
for experimental data of $\Omega$ and $\phi$. The coefficient $M_{q}$
is 1.16, 1.15, 1.18 at three collisional energies, respectively. The
deviation of the $M_{q}$ from 1 is due to the influence of three
parts. Numerical calculations give the $(A_{\Omega}^{(c)\frac{1}{3}}A_{\phi}^{(p)\frac{1}{2}})/(A_{\Omega}^{(p)\frac{1}{3}}A_{\phi}^{(c)\frac{1}{2}})$
is 1.03, 1.02, 1.06, the $(g_{B}^{\frac{1}{3}}(z_{c})g_{M}^{\frac{1}{2}}(z_{p}))/(g_{B}^{\frac{1}{3}}(z_{p})g_{M}^{\frac{1}{2}}(z_{c}))$
is 1.05, 1.04, 1.01, and the $(N_{coll}^{(c)}/N_{q}^{(c)})^{\frac{1}{6}}/(N_{coll}^{(p)}/N_{q}^{(p)})^{\frac{1}{6}}$
is 1.07, 1.08, 1.10, respectively. 

\begin{figure*}
\centering{}\includegraphics[width=0.7\linewidth]{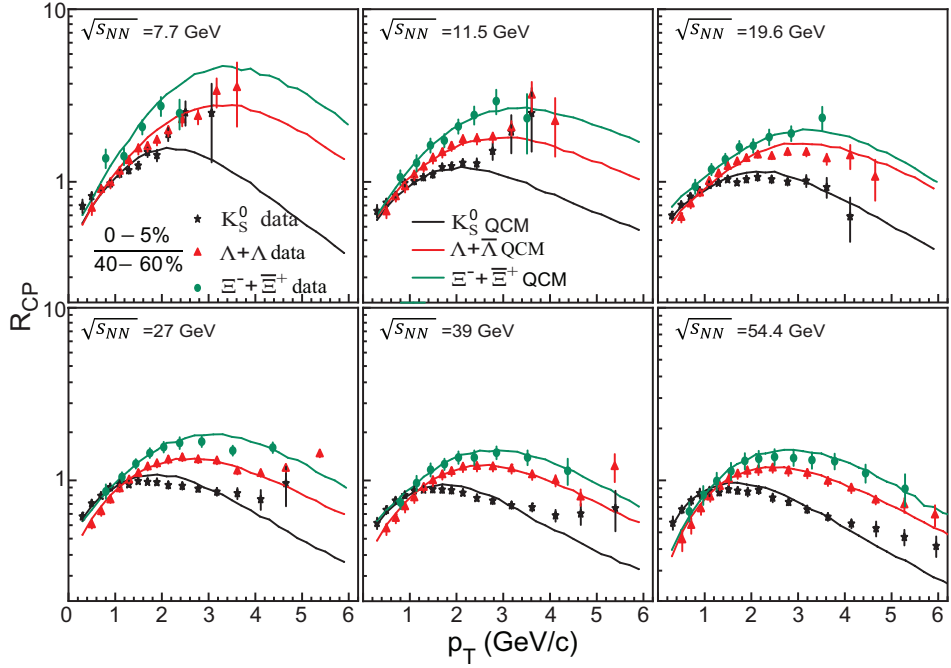}\caption{$R_{CP}$ of $K_{S}^{0}$, $\Lambda+\bar{\Lambda}$ and $\Xi^{-}+\bar{\Xi}^{+}$
(0-5\%)/(40-60\%) at mid-rapidity ($|y|<0.5$) in Au+Au collisions
at $\sqrt{s_{NN}}$=7.7-54.4 GeV. Symbols are experimental data \citep{adam2020strange,Muhammad:2019hbu}
and lines are results of our model.\label{fig:fig19}}
\end{figure*}

$R_{CP}$ of $K_{S}^{0}$ , $\Lambda$ and $\Xi$ in the EVC mechanism
are
\begin{align}
 & R_{CP,\Lambda}(p_{T})\label{eq:RCP_lamb}\\
 & =\frac{A_{\Lambda}^{(c)}g_{B}(z_{c})}{A_{\Lambda}^{(p)}g_{B}(z_{p})}\left(\frac{N_{coll}^{(c)}/N_{q}^{(c)}}{N_{coll}^{(p)}/N_{q}^{(p)}}\right)^{2}R_{CP,u}^{2}(x_{u}p_{T})R_{CP,s}(x_{s}p_{T}),\nonumber 
\end{align}
\begin{align}
 & R_{CP,\Xi}(p_{T})\label{eq:RCP_xi}\\
 & =\frac{A_{\Xi}^{(c)}g_{B}(z_{c})}{A_{\Xi}^{(p)}g_{B}(z_{p})}\left(\frac{N_{coll}^{(c)}/N_{q}^{(c)}}{N_{coll}^{(p)}/N_{q}^{(p)}}\right)^{2}R_{CP,u}(x_{u}p_{T})R_{CP,s}^{2}(x_{s}p_{T}),\nonumber 
\end{align}
\begin{align}
 & R_{CP,K}(p_{T})\label{eq:RCP_K}\\
 & =\frac{A_{K}^{(c)}g_{M}(z_{c})}{A_{K}^{(p)}g_{M}(z_{p})}\left(\frac{N_{coll}^{(c)}/N_{q}^{(c)}}{N_{coll}^{(p)}/N_{q}^{(p)}}\right)R_{CP,u}(x_{u}p_{T})R_{CP,s}(x_{s}p_{T}),\nonumber 
\end{align}
$R_{CP}$ of these hadrons not only depend on the $R_{CP}$ of strange
quarks but also depend on that of up/down quarks. 

Fig. \ref{fig:fig19} shows $R_{CP}$ of $\Lambda$, $\Xi$ and $K_{S}^{0}$
in Au+Au collisions at six collisional energies. Symbols are experimental
data \citep{adam2020strange,Muhammad:2019hbu}. Lines of different
kinds are our model results which are calculated from their model
results of inclusive $p_{T}$ spectra shown in Figs.~\ref{fig:fig4}-\ref{fig:Xibar_fpt}.
We see that the experimental data of three hadrons exhibit some hierarchy
properties. Compared with $R_{CP}$ of $\Lambda$ and $\Xi$, $R_{CP}$
of $K_{S}^{0}$ at $\sqrt{s_{NN}}=19.6-54.4$ GeV reach the maximum
at a smaller $p_{T}$ (i.e., $p_{T}\approx1.5-2$ GeV/c) and is lower
than those of $\Lambda$ and $\Xi$ as $p_{T}\gtrsim2$ GeV/c. $R_{CP}$
of $\Lambda$ is smaller in magnitude than that of $\Xi$ to a certain
extent but is quite similar with the latter in the global $p_{T}$
dependence. Our model reproduce these hierarchy properties. The last
few points of $R_{CP}$ for $K_{S}^{0}$ at $\sqrt{s_{NN}}=$ 7.7
and 11.5 GeV in range $p_{T}\gtrsim2$ GeV/c are close to those of
baryons, which is beyond the model expectation.

Using Eqs.~(\ref{eq:RCP_lamb})-(\ref{eq:RCP_K}) and $R_{CP}$ of
quarks shown in Fig.~\ref{fig:fig18}, we can naturally explain the
experimental data of $R_{CP}$ of $\Lambda$, $\Xi$ and $K_{S}^{0}$.
We see that $R_{CP}$ of $u$, $\bar{u}$ and $s$ are all dependent
on $p_{T}$. $R_{CP}$ of (anti-)quark increases with $p_{T}$ at
low $p_{T}$ and turns to decrease with $p_{T}$ as $p_{T}\gtrsim$
1 GeV/c. $R_{CP}$ of $K_{S}^{0}$ relates to the product of two quark
$R_{CP}$, and therefore $R_{CP}$ of $K_{S}^{0}$ reaches the maximum
at $p_{T}\approx2$ GeV/c. $R_{CP}$ of $\Lambda$ and $\Xi$ relate
to the product of three quark $R_{CP}$, therefore they reaches the
maximum at $p_{T}\approx3$ GeV/c and the maximum values are higher
than that of $K_{S}^{0}$. This is quite similar to the case of $\Omega$
and $\phi$ discussed above. We also see from Fig.~\ref{fig:fig18}
that $R_{CP}$ of strange quarks has a stronger non-monotonic $p_{T}$
dependence than that of $u$ and $\bar{u}$. In addition, because
the fraction of strange quarks (i.e., $\lambda_{s}=N_{s}/N_{\bar{u}}$)
in central collisions is higher than that in peripheral collisions,
$R_{CP}$ of strange quarks is globally higher than that of $u$ or
$\bar{u}$ to a certain extent. Since $\Xi$ has two strange quarks,
$R_{CP}$ of $\Xi$ has a stronger $p_{T}$ dependence than that of
$\Lambda$ and is globally higher than that of $\Lambda$.

\section{Summary\label{sec:Summary}}

In this paper, we have applied an equal-velocity quark combination
to systematically study $p_{T}$ spectra of strange hadrons $K_{S}^{0}$,
$\phi$, $\Lambda$, $\Xi^{-}$, $\Omega^{-}$, $\bar{\Lambda}$,
$\bar{\Xi}^{+}$ and $\bar{\Omega}^{+}$ at mid-rapidity in Au+Au
collisions at $\sqrt{s_{NN}}=$ 7.7, 11.5, 19.6, 27, 39 and 54.4 GeV.
The model was proposed in \citep{Song:2017gcz,Gou:2017foe} by inspiration
of the quark number scaling property of hadronic $p_{T}$ spectra
in $pp$ and $p$Pb collisions at LHC energies and has a series of
successful applications in describing hadron production in $pp$ and
$p$Pb collisions at LHC as well as AA collisions at both RHIC and
LHC \citep{Song:2018tpv,Li:2017zuj,Zhang:2018vyr,Song:2019sez,Song:2020kak,Li:2021nhq,Wang:2019fcg}.
Application of this model to STAR BES energies can further test the
universal property of the hadronization in different collisions. In
the study, we focus on the self-consistent explanation on $p_{T}$
spectra of strange hadrons at STAR BES energies. Therefore we not
only carried out the global comparison with $p_{T}$ spectra data
of these hadrons but also concentrated on baryon-to-meson ratios and
nuclear modification factor in the low and intermediate $p_{T}$ range
which are sensitive to hadronization mechanism. 

We firstly carried out a global fit to experimental data of $p_{T}$
spectra of strange hadrons. The model has three quark inputs, i.e.,
$f_{u}(p_{T})$, $f_{\bar{u}}(p_{T})$ and $f_{s}(p_{T})$. We used
the data of $\Lambda$, $\bar{\Lambda}$ and $\phi$ to fix them and
subsequently calculated $p_{T}$ spectra of $K_{S}^{0}$, $\Xi^{-}$,
$\Omega^{-}$, $\bar{\Xi}^{+}$, $\bar{\Omega}^{+}$ and compared
them with experimental data. We evaluated the relative deviation between
model calculation and experimental data of these eight hadrons. We
found that the relative deviation is generally about 2-3\% at $\sqrt{s_{NN}}=$
27, 39, 54.4 GeV and in central collisions at 7.7, 11.5, 19.6 GeV.
The deviation slightly increases up to about 4\% in semi-central and
peripheral collision at $\sqrt{s_{NN}}=$ 7.7, 11.5, 19.6 GeV. These
results indicate that our model can give a globally consistent explanation
on $p_{T}$ spectra of these strange hadrons at the studied collision
energies. 

We studied the dependence of two baryon-to-meson ratios $\bar{\Lambda}/K_{S}^{0}$
and $\Omega/\phi$ on $p_{T}$, collision centrality and collision
energy. By classifying the property of quark $p_{T}$ spectra in the
range $p_{T}\lesssim1$ GeV/c and that in the range $p_{T}\gtrsim1$
GeV/c, we provided an intuitive explanation on the increase of two
ratios in the range $p_{T}\lesssim2-3$ GeV/c and their subsequent
decrease at larger $p_{T}$. Combining the extracted quark $p_{T}$
spectra at hadronization, we further discussed the quark level origin
of the change of global magnitude and the movement of peak position
of two ratios in different collision centrality and at different collision
energies.

We studied the nuclear modification factor $R_{CP}$ of strange hadrons.
Taking advantage of the analytic feature of EVC mechanism, we derived
the analytic expression for $R_{CP}$ of hadrons and found that $R_{CP}$
of hadrons can be written as the product of those of quarks at hadronization
besides some $p_{T}$-independent coefficients. Using these analytic
formulas, we gave an intuitive explanation on the difference between
$R_{CP}$ of meson and that of baryon, including the difference in
peak position and peak value of the $R_{CP}$. In addition, the hadron
species dependence of $R_{CP}$ of $\Lambda$, $\Xi$, $\Omega$ can
be also naturally understood by considering the property of $R_{CP}$
of quarks at hadronization.

\section{Acknowledgments}

This work is supported in part by the National Natural Science Foundation
of China under Grant No. 11975011 and 12175115, Shandong Provincial
Natural Science Foundation (Grant No. ZR2019YQ06, ZR2019MA053), and
Higher Educational Youth Innovation Science and Technology Program
of Shandong Province (Grant No. 2019KJJ010).

\bibliographystyle{apsrev4-1}
\bibliography{ref}

\end{document}